%% file: jsa_paper.tex
\documentclass[5p]{elsarticle}

\journal{Journal of Systems Architecture}

\usepackage[font={footnotesize},textfont={onehalfspacing}]{caption}
\usepackage{acronym}
\input{acronyms}
\usepackage{url}
\usepackage{csquotes}
\usepackage{amsmath,amssymb,amsfonts}
\usepackage{algorithmic}
\usepackage{pgfplots}
\usepackage{todonotes}
\usepackage{amsthm}

\newtheorem{mydef}{Definition}

\begin{document}

\begin{frontmatter}

\title{Towards a Real-Time IoT: Approaches for Incoming Packet Processing in Cyber-Physical Systems\tnoteref{t1}}
\tnotetext[t1]{This is the preprint of an article to be published in the Journal of Systems Architecture (\url{https://doi.org/10.1016/j.sysarc.2023.102891}). CC BY-NC-ND}

\author[tuberlin]{Ilja Behnke\corref{mycorrespondingauthor}}
\cortext[mycorrespondingauthor]{Corresponding author}
\ead{i.behnke@tu-berlin.de}
\author[tuberlin]{Christoph Blumschein}
\author[tuberlin]{Robert Danicki}
\author[tuberlin]{Philipp Wiesner}
\author[glasgow]{Lauritz Thamsen}
\author[tuberlin]{Odej Kao}

\address[tuberlin]{Technische Universität Berlin, Distributed and Operating Systems, Berlin, Germany}
\address[glasgow]{University of Glasgow, School of Computing Science, Glasgow, United Kingdom}

\begin{abstract}
Embedded real-time devices for monitoring, controlling, and collaboration purposes in cyber-physical systems are now commonly equipped with IP networking capabilities. However, the reception and processing of IP packets generates workloads in unpredictable frequencies as networks are outside of a developer's control and difficult to anticipate, especially when networks are connected to the internet. As of now, embedded network controllers and IP stacks are not designed for real-time capabilities, even when used in real-time environments and operating systems. 

Our work focuses on real-time aware packet reception from open network connections, without a real-time networking infrastructure. This article presents two experimentally evaluated modifications to the IP processing subsystem and embedded network interface controllers of constrained IoT devices. The first, our software approach, introduces early packet classification and priority-aware processing in the network driver. In our experiments this allowed the network subsystem to remain active at a seven-fold increase in network traffic load before disabling the receive interrupts as a last resort. 
The second, our hardware approach, makes changes to the network interface controller, applying interrupt moderation based on real-time priorities to minimize the number of network-generated interrupts. 
Furthermore, this article provides an outlook on how the software and hardware approaches can be combined in a co-designed packet receive architecture.
\end{abstract}


\end{frontmatter}

\graphicspath{{figures/}{./}} %

\input{inc/total}

\bibliography{mybibfile}

\end{document}

%% file: acronyms.tex
\newacro{nic}[NIC]{Network Interface Controller}
\newacro{irq}[IRQ]{Interrupt Request}
\newacro{dma}[DMA]{Direct Memory Access}
\newacro{cpu}[CPU]{Central Processing Unit}
\newacro{isr}[ISR]{Interrupt Service Routine}
\newacro{ist}[IST]{Interrupt Service Task}
\newacro{tcm}[TCM]{Tightly Coupled Memory}
\newacro{uart}[UART]{Universal Asynchronous Receiver/Transmitter}
\newacro{mmu}[MMU]{Memory Management Unit}
\newacro{nop}[NOP]{No Operation}
\newacro{bd}[BD]{Buffer Descriptor}
\newacro{plc}[PLC]{Programmable Logic Controller}
\newacro{mmio}[MMIO]{Memory-Mapped Input/Output}
\newacro{rx}[RX]{Receive}

\newacro{os}[OS]{Operating System}
\newacro{rtos}[RTOS]{Real-Time Operating System}
\newacro{posix}[POSIX]{Portable Operating System Interface}
\newacro{rms}[RMS]{Rate Monotonic Scheduling}
\newacro{wcet}[WCET]{Worst Case Execution Time}
\newacro{tcb}[TCB]{Thread Control Block}
\newacro{ipc}[IPC]{Inter-process Communication}
\newacro{api}[API]{Application Programming Interface}
\newacro{fifo}[FIFO]{First In First Out}

\newacro{osi}[OSI]{Open Systems Interconnection}
\newacro{phy}[PHY]{Physical Layer}
\newacro{mac}[MAC]{Medium Access Control}
\newacro{sctp}[SCTP]{Stream Control Transmission Protocol}
\newacro{quic}[QUIC]{Quick \acs{udp} Internet Connections}
\newacro{mtu}[MTU]{Minimum Transmission Unit}

\newacro{csv}[CSV]{Comma Seperated Values}

\newacro{pc}[PC]{Personal Computer}
\newacro{iot}[IoT]{Internet of Things}
\newacro{dos}[DoS]{Denial of Service}
\newacro{qos}[QoS]{Quality of Service}
\newacro{tsn}[TSN]{Time Sensitive Networking}

\newacro{bsd}[BSD]{Berkely Software Distribution}
\newacro{lwip}[lwIP]{Lightweight IP}
\newacro{lrp}[LRP]{Lazy Receiver Processing}

\newacro{hp}[HP]{High Priority}
\newacro{lp}[LP]{Low Priority}
\newacro{mp}[MP]{Medium Priority}

%% file: inc/total.tex
\section{Introduction}
With the introduction of the Internet of Things (IoT) in manufacturing industries and automotive systems IP networks have entered the domain of real-time embedded systems. Technologies such as 5th generation (5G) mobile networks have been driving the integration of such systems in business networks, command and control infrastructures, and machine-to-machine (M2M) communication \cite{cheng2018industrial, leonardi2019rt, foukalas2019dependable}. Furthermore, commands that are being sent over IP networks are subject to real-time requirements in applications like remote control of industrial machines, remote surgery, and autonomous routing of logistic robots \cite{laaki2019prototyping, liu2020latency, lee2018development}.
To this end, recent research has investigated industrial network architectures in the Industrial IoT (IIoT) and \ac{tsn}~\cite{li2022joint, bruckner_introduction_2019, vitturi2019industrial}. However, the impact of IP networking on the real-time behavior of embedded systems has largely been ignored. 

Embedded real-time devices need to be designed holistically and take hardware cost, energy efficiency and robustness into account, while computing power is typically constrained~\cite{holistic_scheduling_distributed_rt}. At the same time, when devices are used as controlling units in cyber-physical systems, they demand predictable and limited execution times~\cite{alcacer2019scanning, nguyen_real-time_2020}. Input devices like sensors or communication interfaces have always undermined this predictability, as the used \acp{irq} preempt running real-time process regardless of their priorities~\cite{irq_mngmnt_rt}. Yet, short \acp{isr} and controllable IRQ frequencies enabled real-time system developers to incorporate these into worst case execution time analyses, making external interrupts manageable. 

In this regard, \acp{nic} act similarly to other I/O devices. When a network packet is received, the content is written to memory and an IRQ is triggered. The preempting ISR then pre-handles the packet before notifying the operating system. Then, a usually highly prioritized network task resumes packet processing in order of entry and forwards it to a waiting application socket. All these tasks are performed independently of the presumed priority of the packet and create a timing overhead proportional to the rate of incoming packets~\cite{behnke_interrupting_2020, niedermaier2018you}. In common IP networks this rate is not predictable by a real-time applications developer. Yet, IoT use-cases may contain real-time requirements and traditional IP networking without real-time specific protocols or hardware at hand~\cite{tahaei2020rise}. 

Evaluations on common off-the-shelf IoT microcontrollers showed that network controllers and IP stack implementations are not prepared for to real-time environments~\cite{behnke_interrupting_2020}. The tested microcontrollers were overloaded already with 1000 packets per second with no apparent mitigation reactions other than network task and system shut down.
In light of this problem, developers must resort to disabling interrupts during critical executions or require separate resources for networking and processing in resource-constrained microcontroller environments.
However, these solutions are not practical in scenarios where IP networks are used to control cyber-physical systems conveying soft real-time messages that are relevant for baseline functionality. For example, complex industrial actuators with internal feedback control implemented on constrained embedded devices need to meet local real-time requirements independently of the network load while prioritizing the reception of their real-time messages among the received traffic.

While there is extensive research on enabling technologies in the areas of communication and integration in IoT, device architectures have received less attention~\cite{bansal2020iot, alcacer2019scanning, qiu2020edge}. Recent approaches focus on blocking or processing traffic before it arrives at the device~\cite{mandalari2021blocking, niedermaier2019secure, haar2019fane}. Previous works failed to include network-specific factors and focused on interrupt management only~\cite{gomes2015task, multisloth}. Past work has proposed to do packet classification in the networking driver as early as possible~\cite{druschel1996lazy} and identify a priority for each UDP-packet by receiving process and defer the subsequent packet processing~\cite{lee_interrupt_2010, lee_priority-based_2015} based on the assigned priority. Furthermore, solutions exist using specialized networking technologies such as \ac{tsn}~\cite{wollschlaeger_future_2017, schriegel_migration_2021, silva_adequacy_2019} or Software Defined Networking (SDN)~\cite{lin_dte-sdn_2018, henneke_analysis_2016, foschini_sdn-enabled_2021}. While these could be used to mitigate the problem of high packet rates, they require specialized cooperative networking hardware and a rigorous, inflexible real-time network architecture~\cite{sudhakaran_enabling_2021, schriegel_migration_2021}. To the best of our knowledge, there is no published research on extending NICs and network stack implementations to facilitate connected real-time embedded systems in common IP networks.

\subsection*{Contributions}
This is an extended discussion of the work first presented at the International Symposium on Real-Time Distributed Computing 2022 \cite{blumschein2022differentiating} and also includes material previously displayed at the poster session of the Symposium on Applied Computing 2022~\cite{behnke2022priority}.

With this article, we focus on the real-time implications of IP packet reception in real-time systems by proposing two different network subsystem modifications. Both approaches differentiate incoming packets by their priority. The priority of packets is determined by the receiving task priority and real-time requirements in the \ac{rtos}. The approaches regard constrained IoT devices with simple NICs and network stacks, not able to participate in real-time specific networking via Time-Sensitive or Software Defined Networking solutions. The devices communicate using unmodified IP networking. 

The first implementation, a \textbf{software modification}, reduces the impact of best-effort packet processing on the real-time behavior in IoT devices. This is achieved using an early priority-dependent demultiplexing scheme for incoming packets and subsequent aperiodic per-flow scheduling in the network driver. The modification protects real-time embedded systems against network-induced system overloads while optimizing for low-latency processing of high-priority IP flows. This is achieved by strictly controlling the best-effort performance of low-priority flows.

The second approach addresses the problem with a \textbf{hardware modification} of network interface controllers. By mapping IP flows to the priorities of receiver processes, the priority space of an \ac{rtos} can be extended to include the moderation of network-generated interrupts. By applying different interrupt moderation parameters of priority-based receive queues and dropping unregistered flows, the amount of interrupts is reduced while high-priority packets are still handled immediately.

Both presented approaches include experimental evaluations under varying network loads on prototypical implementations including real hardware and simulations. 

Lastly, we present an outlook on how our hardware and software approaches can be combined effectively to a \textbf{unified design}. 

\subsection*{Outline}
Section \ref{sec:background} provides the relevant background information.
Section \ref{sec:related_work} presents the related work.
Section \ref{sec:considerations} presents the preliminary considerations made for the two approaches.
Section \ref{sec:ipstack} presents the software-based modification of the IP subsystem and \ref{sec:ipstack_eval} discusses its evaluation.
Section \ref{sec:nic} and \ref{sec:nic_eval} do the same for the hardware-based adaptation of a NIC.
Section \ref{sec:outlook} gives an outlook towards a combined packet receive architecture.
Section \ref{sec:conclusion} concludes this article.

\section{Background}
\label{sec:background}
This section gives a short introduction to the reception of network packets in \acp{rtos}, interrupt handling, and interrupt moderation.

\subsection{Direct Memory Access}
\label{sec:dma}
In order to relieve the CPU from actively pulling data from or pushing data to the NIC memory, \ac{dma} has been established as a common feature in peripheral hardware. With it, the NIC can asynchronously place received packet data into main memory at previously assigned locations and afterwards only needs to notify the CPU about the arrival of new packets via interrupt. 

\begin{figure}
	\centering
	\includegraphics[scale=0.45]{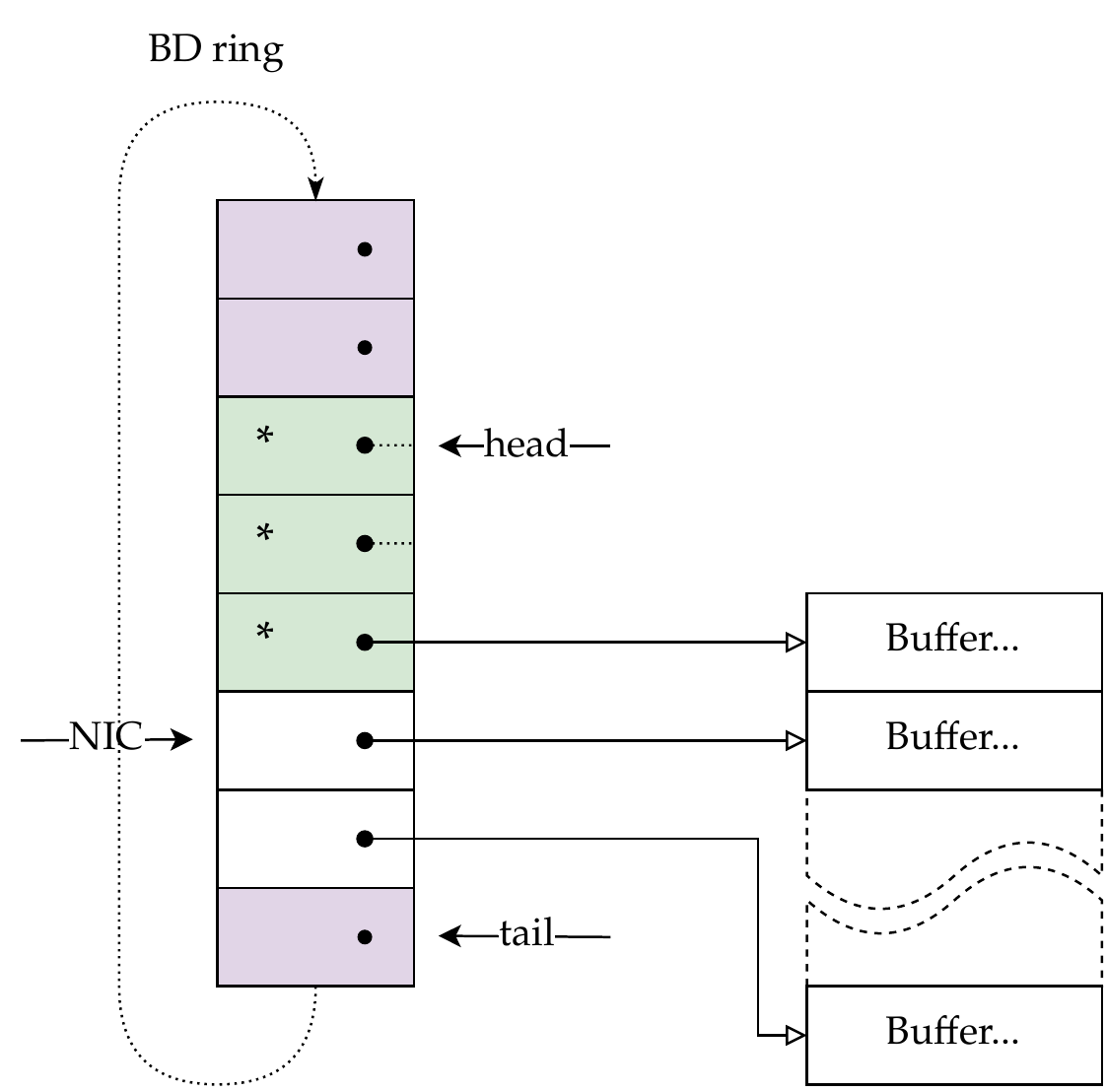}
	\caption{\textbf{BD Ring:} The data structure allows simultaneous access from the NIC adding entries and the CPU checking off processed descriptors.}
	\label{fig:BD_ring}
\end{figure}

The memory assignment usually happens in so-called \ac{bd} rings, as outlined in Figure~\ref{fig:BD_ring}. In memory such a ring comprises an array of Buffer Descriptors, interpreted as a ring buffer. Each \ac{bd} contains a memory pointer to the respective buffer and some metadata for cooperation. The latter typically includes an ownership bit, indicating whether the CPU or NIC is obligated to go on with processing, and a length field indicating how far into the buffer data shall be sent or has been received respectively. This way, both actors can track their current working position(s) individually.

\subsection{Receive Path in IP Networking}
\label{sec:rx_path}
\begin{figure}[h]
    \centering
	\includegraphics[width=\columnwidth]{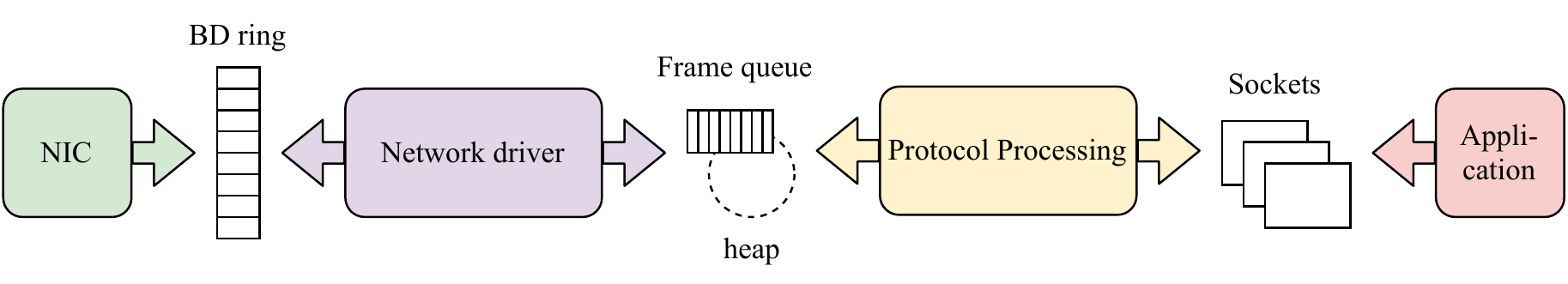}
	\caption{\textbf{\ac{rx}-Path:} Packets are handled by OS data structures, the network driver, and the networking stack implementation before they can be accessed by the receiving application.}
	\label{fig:rx_path_overview}
\end{figure}

At a high level, the \ac{rx} path is organized into subsequently executed stages as seen in Figure \ref{fig:rx_path_overview}.
Upon packet reception, the \ac{nic} transfers the packet content to a previously prepared memory location via \ac{dma}, marks the corresponding \ac{bd} entry and triggers an interrupt.
The network driver, handling the interrupt, acknowledges the \ac{dma}-operation and exchanges the received frame buffer with a newly allocated one.
From here, protocol processing can commence disregarding the already finished MAC-layer operations.

The different network stack implementations vary in their set of features. The \textsc{lw}IP network stack is widely spread among embedded applications and sets its focus on memory efficiency~\cite{lwip}. It covers the majority of commonly used and necessary protocols from Layer 2 up to Layer 4 like ARP, IP, TCP, UDP, DNS, and DHCP. It offers different APIs for efficient access, multi-threading, and an implementation of the Berkeley Socket API. Internally, \textsc{lw}IP does not represent a complete data frame but stores a subset of data in \texttt{pbuf} structures. Other stack implementations like FreeRTOS+TCP\footnote{\url{https://www.freertos.org/FreeRTOS-Plus/FreeRTOS_Plus_TCP/}} offer the complete frame and stick to the Berkeley sockets API while being thread-safe. Using sockets, application tasks can register to receive and transmit through their desired ports.

\subsection{Interrupt Scheduling}
\label{sec:irq}

\ac{irq} and task priorities form two independent priority spaces. Hence, hardware interrupts, such as those triggered by incoming network packets, introduce some challenges to scheduling as they might take over CPU resources at any time.
Yet, when systematically tamed to known minimum inter-arrival times and \acp{wcet}, an integration into the considerations of a schedulable task set becomes possible.
More problematic is the triggered execution of \acp{isr} in an elevated \ac{irq}-context.
It may be either completely uninterruptible itself, or only by another higher priority \ac{irq}-source ("interrupt nesting").

To minimize the worst-case latency incurred by priority inversion situations between interrupts and high-priority tasks, a widespread programming best practice is to reduce the work done in an \ac{isr} to a minimum, only unblocking a deferred \ac{ist} that then does the actual processing.
This compromises on interrupt handling performance for better scheduler control.
Therefore it is often weighed by the driver developer how much additional latency is acceptable until an \ac{isr}/\ac{ist} split is introduced.

\subsection{Aperiodic Scheduling}
\label{sec:aperiodic}
Received network packets generate workload that can be characterized as an aperiodic task inside the \ac{rtos}. 
One approach to integrate aperiodic tasks into fixed-priority scheduling uses so-called server tasks~\cite{sprunt1989aperiodic}.
To the scheduler these behave as ordinary prioritized tasks.
Opposed to other tasks they have no individual objective.
Instead, they use their budget to serve the execution of aperiodic jobs.
Due to their limited budget in each period they can be easily included into scheduling considerations.

A very simple yet effective aperiodic events server is the deferrable server~\cite{strosnider1995deferrable}, which we also utilize in our real-time receive architecture.
It has a limited CPU-time budget to serve aperiodic events.
When the budget is depleted, it pauses execution.
At the end of each period, the server budget gets restored to the initial amount.
A big advantage is the simplicity of the mechanism and therefore of an implementation for that server scheme.
Yet, the deferral of budget consumption incurs on a higher worst-case processing demand than one budget per period, due to possible back-to-back execution patterns.

Let $p$ be the server period and $e$ its execution budget for a period.
There may arrive jobs just before the end of a period consuming the whole capacity $e$ of the server for this period.
With the start of the next period and the consequent budget replenishment, another duration $e$ may be serviced to jobs.
In the worst case, we need to expect one extra execution budget. Thus, the highest possible server demand $d(\Delta)$ inside an arbitrary interval can be indicated as
\[ d(\Delta) = e \cdot \left(\left\lceil \frac{\Delta}{p} \right\rceil + 1\right) \]
Note that at least for a small period $p$, the CPU bandwidth $\frac{d(\Delta)}{\Delta}$ still approaches the theoretical server optimum $\frac{e}{p}$.

\subsection{Interrupt Moderation}
\label{sec:interrupt_moderation}
To decrease the performance impact of incoming packets, high performance NICs employ interrupt moderation techniques. Instead of sending an interrupt for each received data frame, the NIC delays the delivery of an interrupt in order to receive and coalesce additional packets~\cite{makineni2006receive}. Different strategies to realize the delay exist. 

A simple approach is to use a \emph{packet counter} that triggers an interrupt and resets once a certain number of packets have arrived. This leads to a constant and homogeneous reduction of interrupts but also introduces the possibility of starving packets and very unpredictable packet delays. To have control over the time packets reside in memory unnoticed, different types of delay timers are applied: 

The \emph{absolute timer} begins a countdown once a packet has been received and only triggers an interrupt once reaching zero. All packets received in this time frame are announced by this interrupt and do not reset the timer. The obvious disadvantage of this approach is the high latency the first packet of each countdown experiences. In low traffic scenarios, this is highly inefficient. 

To this end, \emph{packet timers} can be introduced. Instead of having a relatively long countdown timer to trigger an interrupt for multiple packets, the counter is a lot smaller and resets with each incoming packet. In low traffic scenarios this leads to smaller delays while interrupts can be entirely impeded under high traffic. The mostly applied solution is therefore a combination of a longer absolute timer and a shorter packet timer. The tuning of the specific parameters is highly dependent on the expected load and subject to research in high performance computing \cite{interruptmoderation_hpc}.

\section{Related Work}
\label{sec:related_work}
The introduction of unpredictability in real-time environments through interrupts has been a long-standing research topic. In the following, we present past approaches to mitigate interrupt impact as well as approaches towards real-time aware network processing.

\subsection{Interrupt Management in Real-Time Systems}
The Advanced Interrupt Controller~\cite{gomes2015task} monitors the priority of the currently running process to determine if an interrupt should be triggered or held back by comparing it to the interrupt priority. A simple extension of the interrupt controller unifies the priority spaces of attached interrupts and operating system processes. However, this does not facilitate for the circumstances around network packets since interrupt priorities of all packets are the same and different packets cannot trigger interrupts of different priorities. 

Prominent work regarding the unification of priority spaces is the approach implemented in the \textit{Sloth} OS \cite{sloth} \cite{sleepy_sloth}. Sloth implements a general abstraction for software threads and ISR, abolishing their distinction. Instead of using a software scheduler for threads, every control flow is designed as a thread-related system call using the hardware interrupt system. By letting the hardware manage all control flows, context switches have less overhead and --- which is more interesting here --- ISR and (other) threads preempt each other in accordance with their priorities. While this abolishes the problem of priority inversion, high packet loads still lead to high interrupt frequencies, impacting real-time tasks.

The priority inversion impact of interrupts in real-time systems has been identified and tackled by Amiri et. al. by employing priority inheritance protocols for interrupt service threads~\cite{amiri2015predictable}. This approach however only works for the schedulable part of interrupt handling of device drivers. 

Using interrupt moderation to relieve the CPU in high traffic scenarios is a method studied mainly for high throughput devices, as systems connected to Gigabit Ethernet networks are subject to potentially millions of packets per second~\cite{gebert2016performance}. However, some work also exists studying embedded devices running the Linux kernel. Spanos et al. evaluate the performance implications of advanced interrupt handling techniques in the "New API" Linux device driver extension~\cite{spanos2008internals}. 

The issue of DoS attacks in industrial IoT environments has been addressed by Niedermaier et al.~\cite{niedermaier2019secure}. A dual microcontroller architecture is proposed to separate networking tasks from critical real-time processes. The presented setup highlights the disproportionate processing requirements of IP networking on microcontrollers. While this does mitigate the effects of DoS attacks, it is not obvious how time critical packets are separated from attack packets. 

The PIERES tool~\cite{bender2021pieres} is a small framework running on real-time IoT devices, designed to analyze the real-time behavior under different network loads and hardware configurations. This playground allows developers and researchers to perform network interrupt experiments on real-time embedded systems with different network interface controller implementations, load generators and timing utilities.

\subsection{Real-Time Aware Packet Processing}
The network stack architecture \emph{Lazy Receiver Processing} (LRP) introduced an important and much used approach still interesting today~\cite{druschel1996lazy}. It improves performance, stability and fairness on server systems with high incoming network throughput. The processing of newly arriving packets can cause persistent crowding of the application processes so that they are not able to receive data. The packets then have to be discarded while the applications continue to starve. 
This situation is prevented by early multiplexing of incoming packets, prioritized execution of the rest of the protocol stack in the context of the receiving process, and thus enabling early discarding of packets on congested paths. This approach would benefit from a pre-sorting of packets by the hardware and the general interrupt decrease attained from our multiqueue NIC. The decrease in throughput as the packet rate increases can thus be prevented or mitigated. By consistently differentiating different network flows, they take an elegant approach that may also efficiently improve real-time behavior in IoT devices.

Building atop the idea of LRP, Lee et al. investigated on reducing the impact of \ac{lp}-packets on the real-time behavior of a network-independent task by introducing port-based prioritization of protocol processing \cite{lee_interrupt_2010, lee_priority-based_2015}.
In order to achieve this, they classify a UDP-packet by it's port in a \enquote{top half} interrupt handler.
By looking up a special port-priority-table, whose data is sourced by all bound UDP-sockets, their top half infers a priority for each packet.
In Linux's \texttt{softirq} scheduling entity belonging to the kernel, they introduce a gate functionality:
Packets are only ever processed as long as their priority is higher than the current active priority of the system.
Otherwise their processing is delayed until at least the next regular \texttt{softirq} invocation.
They consequently show how their modification is able to reduce a long-running critical task lateness measurably.
However their implementation is restricted by the inappropriate scheduling behavior of the \texttt{softirq}-handler in Linux, which is not preemptable even by the most critical processes and gets rescheduled in similar way as polling, adding unnecessary high network latency once packets aren't processed eagerly anymore.
Moreover, their work only considers UDP-packets.
Finally, when considering overloading scenarios a mere flow differentiation and prioritization is not sufficient for protecting execution guarantees, since packets may also arrive at a highly prioritized task port in high quantities.

The time-predictable IP stack tpIP~\cite{schoeberl_tpip_2018} addresses the challenge of real-time communication in cyber-physical systems. To enable timing predictability and \ac{wcet} analysis the proposed stack uses polling functions in the socket API with non-blocking read and write operations. While focusing on timing analysis and predictability, no measures are taken towards processing performance, interrupt scheduling or the issue of traffic overloads.

Strategies presented in~\cite{danicki2021detecting} deal with the detection and mitigation of network packet overloads in real-time systems.
The \emph{Burst Mitigation} approach, limits the amount of \ac{irq}s that may get processed in a time slice, effectively applying a deferrable server scheduling scheme which considers each \ac{irq} a standard-sized job.
While the work does not consider differentiating mitigation measures over different packet flows, the evaluation already hints the practicality of simple mitigation techniques that can be used beneficially in our approach.
The \emph{Queue mitigation} enables back-propagation of frame queue stagnation to the \ac{irq}, making it adopt the priority of the network task with some delay.
This has great effect when the network task is only equipped with medium priority and no particular small scheduling latency is required.
Two other approaches deal with dynamic schemes, taking advantage of the critical task slack time. The evaluation indicates the effectiveness of the static mitigation approaches. The dynamic ones however strongly depend on the cooperativeness of the critical task and it's execution time consistency.

The Linux kernel provides advanced networking capabilities for routing and traffic control (tc)~\cite{hubert2002linux}. The latter provides mechanisms to control IP traffic such as traffic shaping and forwarding. Using queuing disciplines (qdiscs), incoming and outgoing packets can be queued for each network device. Traffic rates can be moderated and packets can be assigned priorities on the basis of meta information.  Similar to the approaches presented in this article, packets can be matched to certain descriptions and inserted to a qdisc depending on the packet classification. Recent work also shows that the novel Linux kernel features eBPF and EDT can be used to improve the scalability of filters in Linux-based networking~\cite{becker2022network}. Due to the packet-proportional overhead, \acp{rtos} do not commonly provide similar tools.

\subsection{Specialized Hardware}
\label{sec:related_nic}
Network Interface Controllers with multiple transmit and receive queues have been introduced by Intel as early as 2007. The goal is to make use of multicore systems by parallelizing network load on the different queues. The trend is to increase the number of queues to facilitate cloud computing as Zhu et. al. showed in 2020~\cite{zhu2020data}. Multiqueueing in general exists for different high-throughput I/O devices but to our best knowledge is not a common sight among real-time or embedded architectures.

Some modern NICs support serving received packets into multiple \ac{bd} rings \cite{nics}. For once, this is useful to distribute packet reception work over multiple CPU cores in high-performance scenarios. The assignment of packets to these queues can also result from a hardware classification based on packet header fields, which may be configurable by the driver~\cite{shinde2013unicorn, tripathi2009crossbow}. However, this is a feature only found in advanced NIC hardware~\cite{honda2014userlevelstacks}.

Loom~\cite{stephens2019loom} is a multiqueue NIC design that moves per-flow scheduling decisions from the software network stack into the NIC. This way, high throughput and homogeneous policy enforcement can be guaranteed while also providing isolation in multi-tenant cloud data centers.

In \cite{lonardo2015fpga} Lonardo et al. present an application specific NIC design run on FPGAs for high energy physics experiments. The design allows a remote \ac{dma} to CPU and GPU memories, relieving the OS from data transfer management, allowing real-time processing on data received over the network. 

\subsection{Industrial Networking}
Low latency, predictability and reliability are longstanding requirements in industry automation. In the 2000s field bus technologies were moved to work via Ethernet connections. Industrial Ethernet encompasses all usage of Ethernet in an industrial setting.
The target applications typically have both latency and reliability requirements and this drives the design of protocols away from traditional Ethernet approaches for collision detection and avoidance.
The most common traditional real-time Ethernet protocols are EtherNet/IP~\cite{brooks2001ethernet}, PROFINET and EtherCAT~\cite{jansen2004real}.
EtherNet/IP uses the Common Industrial Protocol (CIP) over Ethernet to controlling real-time devices on the network.
PROFINET prioritizes traffic on one physical Ethernet network based on real-time classes. Particularly, the PROFINET-IRT class caters for stringent real-time requirements, leveraging globally synchronized time-triggered switches and interfaces and supporting sub-millisecond cycles.
EtherCAT is particularly relevant in industrial control and automation due to its speed and determinism. It is specified to offer cycle times of less than $100\mu s$ by bypassing transfer processing on the slave device and traffic prioritization. 

In 2011, the first part of what was to be known as Audio/Video Bridging (AVB) was published in a series of IEEE Standards for switched Ethernet~\cite{ieee:802.1ba_2021, ieee:802.1Qat, ieee:802.1Qbv, ieee802.1as_2020}.  
With AVB, later renamed to \ac{tsn} a network can be configured to provide \ac{qos} guarantees that allow real-time traffic to flow through the network with extremely low packet loss and with predictable and bounded latency.
The need for \ac{tsn} has been reinforced by Industry 4.0 requirements. Several standards exist to guarantee time synchronization, packet delivery and maximum latency in industrial environments where contracts can be made between senders and receiver~\cite{finn2018introduction}. Approaches to minimize interrupt loads are not part of proposed \ac{tsn} protocols.

In this work, we assume the absence of an industrial networking infrastructure. Hence, aperiodic and unexpected bursts of traffic are possible. Compared to the referred works making similar assumptions, there are none that address the vertical discrimination of packet handling based on flow priorities. The following sections will describe and evaluate such approaches.

\section{Preliminary Considerations}
\label{sec:considerations}
This section highlights considerations concerning the environment, devices, and data flow for both presented approaches.

For the purposes of this article we consider an IoT device to be an embedded system running an \ac{rtos} and a small number of processes with fixed priorities managed by a preemptive scheduler. The embedded device is a constrained IoT node in an IP network serving real-time applications as well as best-effort networking. No specialized real-time network protocols or infrastructure are available to the device.
In the network stack, a driver controls \ac{dma} transfers, establishes cache coherency and passes packet buffers to a networking task by means of a queue.
The device contains a baseline embedded \ac{nic} used to connect to an IP network which itself might be connected to the Internet. 

The proposed approaches make use of IP flow information for real-time aware packet prioritization. Packets are received by real-time processes on the embedded system.

\begin{mydef}
\label{def:1}
Let $\mathcal{F}$ be the set of IP flows received by the regarded system. An IP flow $f  \in \mathcal{F}$ is a sequence of IP packets arriving at the regarded device and for the purposes of this article characterized by the tuple $(\text{Src}, \text{Dst}_\text{port}, P, t_P )$.
\begin{itemize}
\item $\text{Src}$ is the source node, identified by its IP address.
\item $\text{Dst}_\text{port}$ is the destination port number.
\item $P$ is the priority of the flow as observed by the regarded device.
\item $t_P$ is the minimum expected period of the flow, meaning the interarrival time between two packets in the flow. 
\end{itemize}
\end{mydef}

The priority of a flow is a parameter assigned by the receiving system, as the real-time implications to this system need to be considered. The period of a flow depends on the sender, application, and network infrastructure. 

\begin{figure}
\centering
\includegraphics[scale=0.7]{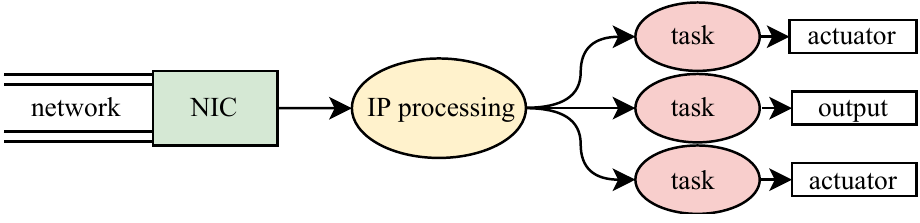}
\caption{\textbf{Data flow:} Path of received messages through the regarded device.}
\label{fig:dataflow}
\end{figure}

\paragraph*{Data Flow}
Each task listens on a separate socket for messages. Each message is delivered using one IP packet. The tasks process the message and produce an output or control a physical actuator. Figure~\ref{fig:dataflow} shows the resulting data flow model.

\subsection{Priority Inversion in the RX Path}
We observe a problem of incorrect priority enforcement in the networking subsystem receive path. Generally, an incoming packet is used by (at most) one specific task. Yet, as long as its purpose is unknown, we have to schedule the processing of each packet equally. For the practical implementation of the networking subsystem in an RTOS, that means assigning a fixed priority to the protocol processing server task, inevitably creating a priority inversion situation:

\begin{itemize}
    \item Using a high priority, as usually found in current embedded frameworks, high priority tasks may starve in favor of low priority packets.
    \item Using a low priority, a receiving high priority task may wait for the end of execution of a medium priority tasks. %
\end{itemize}

The above formulated priority inversion cannot be completely eliminated by its inherent nature. On the one hand, we do not want to spend computing resources until knowing whether it will be worth it considering the current scheduling situation. On the other hand, incorporating the use of a particular packet requires prior protocol handling.

As already shown in~\cite{danicki2021detecting, behnke_interrupting_2020, niedermaier_you_2018}, a first angle to prevent receive packet induced overloads is the introduction of budget enforcement to the entire RX-path. However, when limiting the reception of incoming packets, one might include soft real-time traffic that, while not being critical to the integrity of the controlled cyber-physical system, is relevant for baseline functionalities. This converts the catastrophic situation of potential system failure into a tenable but still undesirable situation of reduced network communication liveness, trading availability for robustness.
Note that we are expecting multiple IP flows with different levels of timeliness requirements and packet frequencies enabling a different grasp on the problem.

\section{The Software Approach: Priority-Aware Scheduling of Incoming Packets}
\label{sec:ipstack}
The first of the two presented approaches is entirely software based. By adapting common light weight IP stacks, we aim to introduce real-time aware packet processing and mitigate the unpredictability of IP networking.

\subsection{Requirements}
A modification of the packet reception subsystem for real-time IoT devices should satisfy the following requirements.

\begin{enumerate}
\item[1.1] \emph{Priority Inversion Mitigation.} Packets of high-priority flows must be processed and delivered before those of low-priority flows. 
\item[1.2] \emph{Overload Protection.} Packet floods must not lead to system overloads. In case of a packet flood, low-priority flows must be  throttled in favor of high-priority processes. A general rate limitation must protect the real-time properties of running tasks.
\item[1.3] \emph{Performance Retention.} The approach should not introduce a longer packet processing delay. However, predictability and overload protection must be prioritized.
\end{enumerate}

\subsection{Overview}
\begin{figure}[h]
    \centering
    \includegraphics[width=\columnwidth]{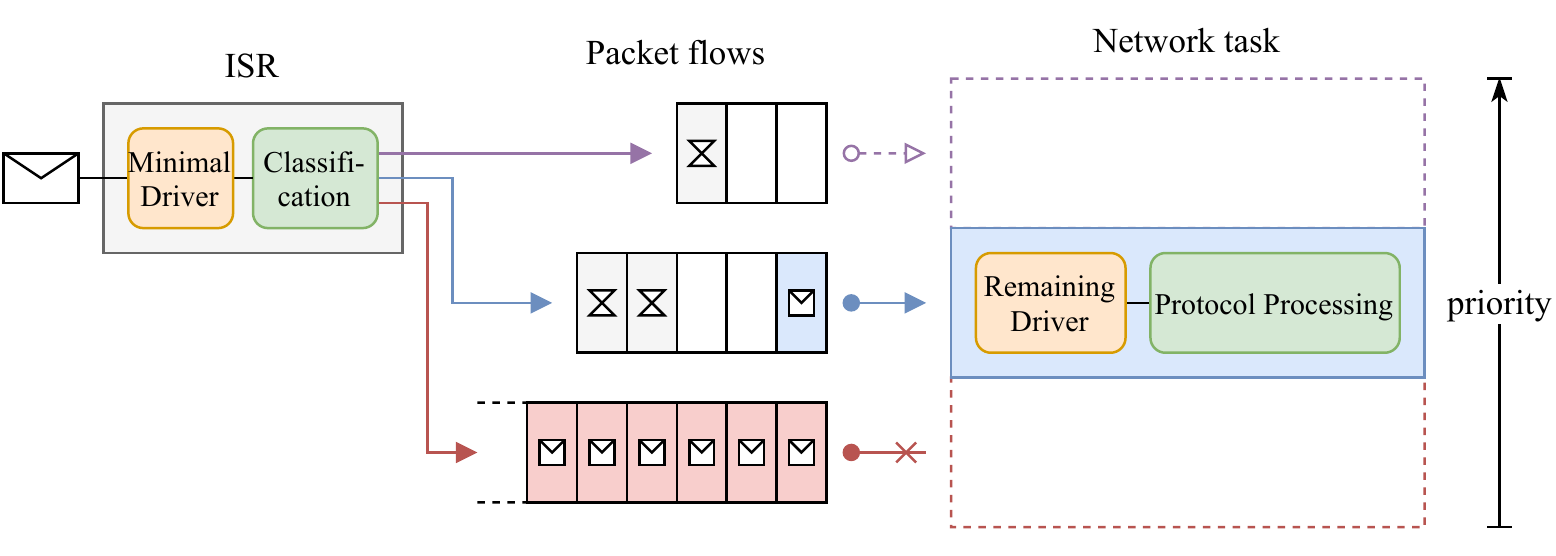}
    \caption{\textbf{Architecture overview:} 
   Packets are classified early and enqueued by their flow, with individual periodic capacity restrictions applied. Further processing is scheduled by inherited priority.}
    \label{fig:approach_overview}
\end{figure}

The proposed architecture is designed around a data structure of differentiated flow queues, which replaces the simple frame queue.
Each flow defines a priority and a period, to affect the further processing of its packets.
For the prioritization of processing, we add a priority manipulation mechanism to the protocol processing task.
In order to gain a maximal advantage from scheduling the subsequent processing stage, the driver is modified to do only the necessary work of classifying incoming packets to flows by their header entries.
The remaining activity is then executed on packet retrieval by the scheduled protocol processing task as presented in Figure~\ref{fig:approach_overview}.

The proposed architecture combines three concepts:

\begin{enumerate}
    \item \emph{Soft Early Demultiplexing} into receiver-centric flows.
    \item \emph{Prioritized Protocol Handling} based on these flows.
    \item \emph{Rate Limitation} applied per flow as well as overall.
\end{enumerate}

While theses are not novel ideas independently, we argue that only in this combination they exhibit properties making for a viable solution to the discussed problem:
\begin{itemize}
    \item \emph{Early Demultiplexing} is necessary for differentiating flows on an End-to-End basis, without reliance on network \ac{qos} and as a result satisfy \emph{Requirement 1.1}.
    \item Proper \emph{prioritization} facilitates best-effort communication processes that utilize background resources on the same system and is necessary to satisfy \emph{Requirements 1.2 and 1.1}.
    \item \emph{Rate limitation} as a last resort protects the system from being vulnerable to unexpectedly high traffic in \ac{hp} flows satisfying \emph{Requirement 1.2}.
\end{itemize}

Thus, being able to fully defend the considered system against flooding induced overload, while at the same time ensuring high connectivity for particular well-behaved \ac{hp}-flows even in scenarios with overall high incoming traffic, and handling \ac{lp}-flows with best-effort resources, this combination forms an IoT real-time aware overload protection.

In the following subsections we introduce the concept of each of the three basic building blocks of our architecture and discuss relevant implementation aspects.

\subsection{Soft Early Demultiplexing}

In order to minimize the effort spent until after classification, we employ Early Demultiplexing \cite{druschel1996lazy}.
By peeking into key header entries, a packet is assigned to its eventual receiver process.

The benefit of demultiplexing performed in software depends heavily on the amount of work that can be saved by mere demultiplexing compared to full protocol processing.
Since the packet scheduling in our architecture can only influence the processing that follows after Early Demultiplexing, the achievable degree of partial network liveness in overload scenarios depends on its quick execution.

\begin{figure}
    \centering
    \includegraphics[width=0.9\columnwidth]{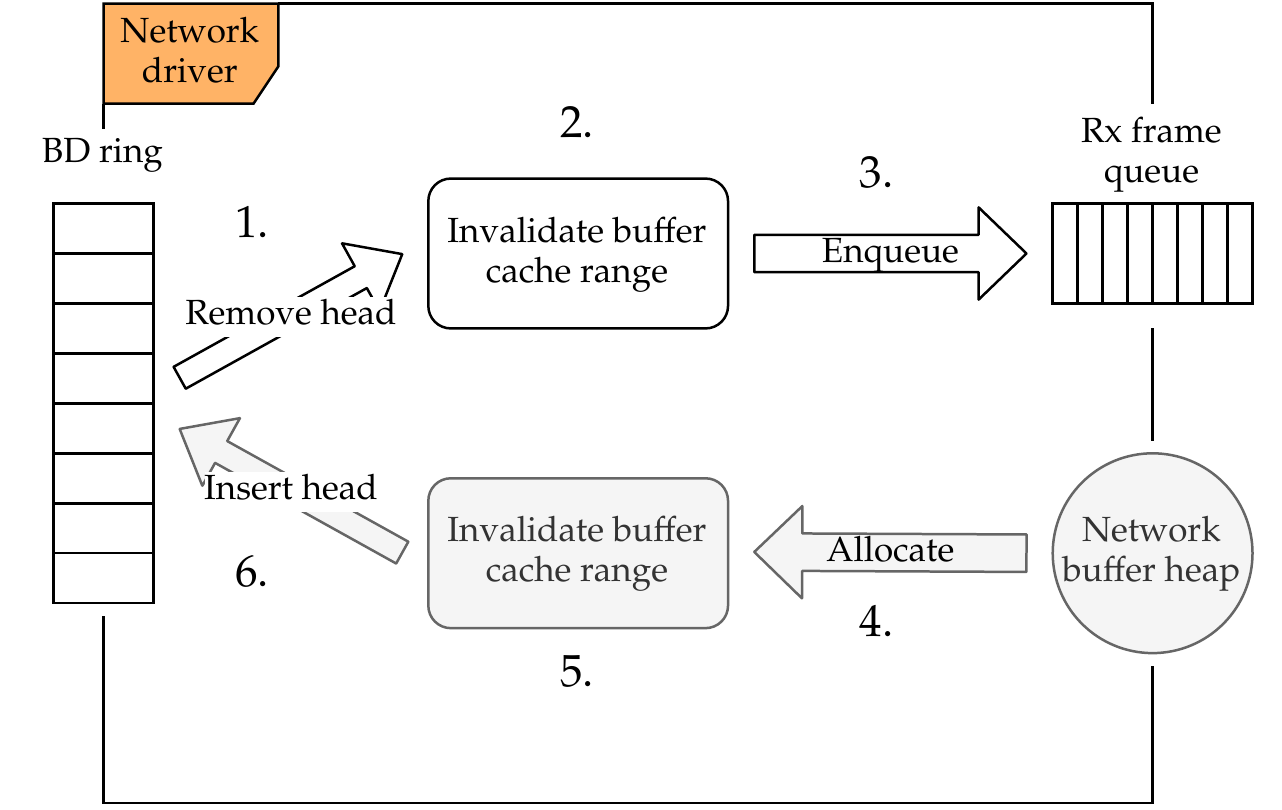}
    \caption[Receive activity in the original driver]{\textbf{Receive activity in the original driver:} 
    Once an interrupt occurs, the driver code moves a packet buffer from the \ac{bd} ring to a simple queue, and fills the vacant position with a newly allocated one. Due to cache coherency requirements of the memory system, the buffer caches have to be invalidated for both the retrieved and the replacement buffer.}
    \label{fig:driver_buffer_flow}
    
\end{figure}

Starting from the existing driver receive path, depicted in Figure \ref{fig:driver_buffer_flow}, we introduce two changes: Packet classification and lazy cache invalidation.

\subsubsection{Packet Classification}

The classification differentiates incoming packets into flows defined by the protocols ARP, ICMP, TCP and UDP.
While the former two form a single flow, the latter are further differentiated by local port numbers in order to implement the receiver task association.

Depending on the used network stack, the lookup from the port to a flow may either be performed using the existent network stack's list of bound socket control blocks, or else requires an additional data structure managed by the driver.
In our prototype based on FreeRTOS+TCP, the socket managing code in the original network stack can easily be locked in a critical section, leaving the \ac{isr} safe to access it.

If a scenario requires anticipating a large number of bound sockets, a sophisticated data structure with better complexity should be employed.
However, with only a few sockets bound at any particular point in time, a linear linked list lookup as found in typical embedded network stacks suffices.

\begin{figure}
    \centering
    \includegraphics[width=0.9\columnwidth]{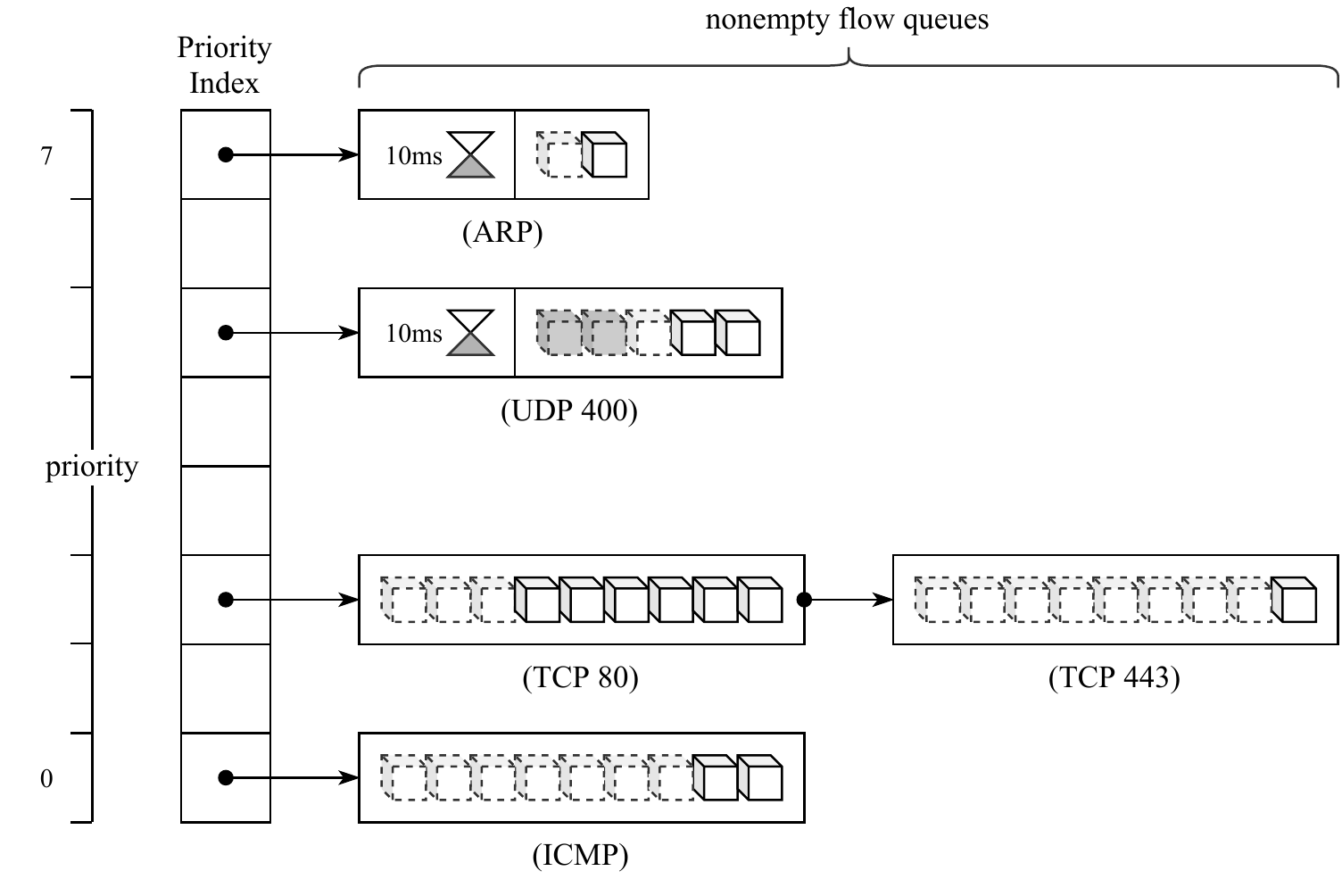}
	\caption[Differentiated flow queues]{\textbf{Differentiated flow queues:}
	Between reception by the driver and further driver and protocol processing, packets get stored in a queue according to their identified flow. These queues are organized by the flow priority, to facilitate fast retrieval of the highest/lowest prioritized packet.
	}
	\label{fig:pdest_depq}
\end{figure}

Instead of enqueueing every received packet to the same \ac{rx} frame queue, each packet is inserted into a specific queue according to the result of the classification.
Because the packets do not necessarily get processed in bounded time, the network subsystem might experience buffer starvation.
To avoid this, buffers of low priority packets are \emph{recycled} when the buffer memory reaches its limit. Buffer recycling here means that the allocated buffer element is freed and a new empty element is appended to the \ac{bd} ring. The packet is effectively dropped.
To this end, the differentiated flow queues are organized in a priority queue structure as depicted in Figure~\ref{fig:pdest_depq}.

The priority of a flow is defined by its respective receiver task and the overall priority space is equal to the one used by the \ac{rtos} task scheduler. This way, packet processing priorities are inherited according to the real-time considerations made for the running processes starting with the enqueueing of packets. Another feasible option is the utilization of network priorities such as the Differentiated Services Field~\footnote{RFC 2474}, flow priorities from real-time protocols, or traffic classes inside TSN-based networks. Yet, we assume environments that do not necessarily provide a network priority, thus relying on receiver priorities.

\subsubsection{Lazy Cache Invalidation}

\begin{figure*}[h]
    \centering
    \includegraphics[scale=0.6]{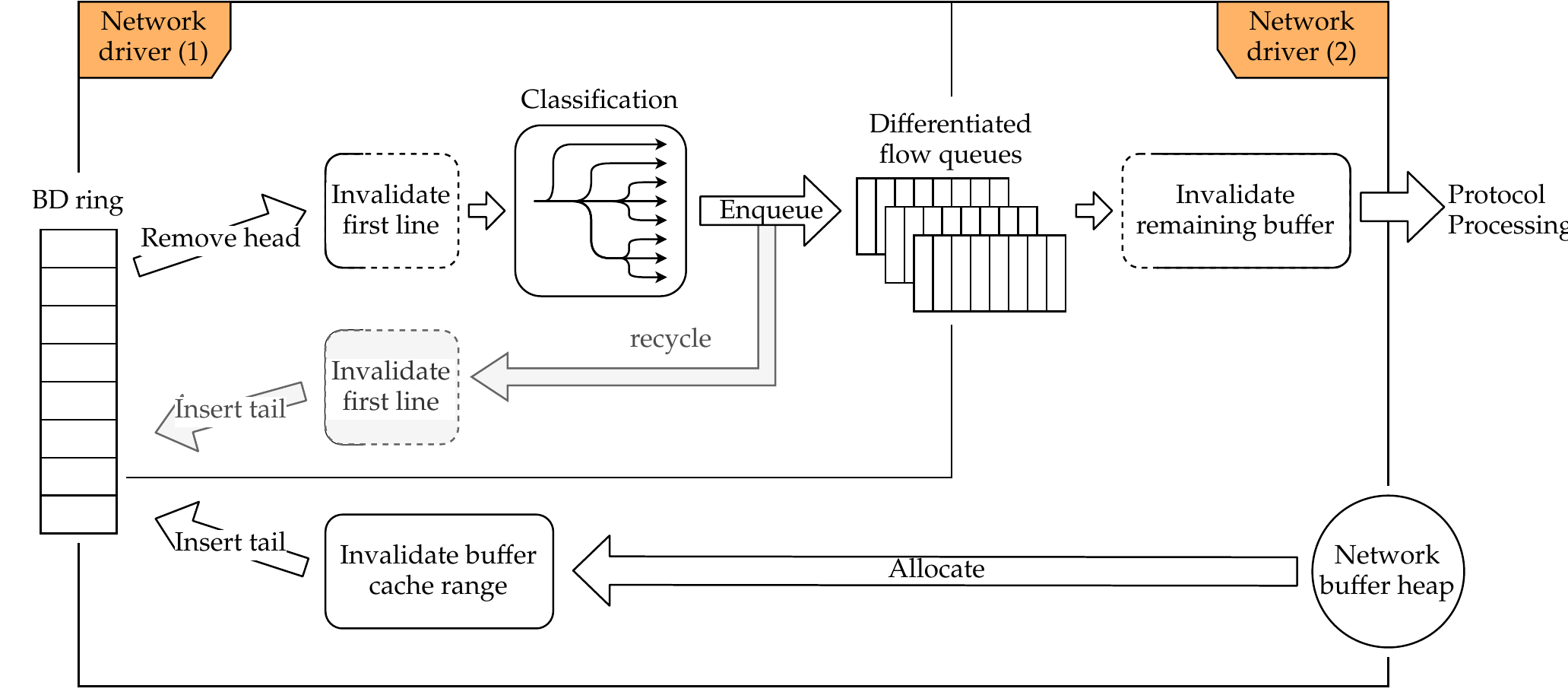}
    \caption{\textbf{Receive driver activity in our approach: }
    The driver is separated into two halves. In the \emph{eager driver} (1), a minimal effort is taken to classify each packet into a flow. As part of the scheduled subsequent protocol processing, the \emph{deferred driver} (2) establishes cache coherency and refills the \ac{bd} ring once the packet is needed.}
    \label{fig:my_driver_buffer_flow_with_recycling}
\end{figure*}

On embedded systems that feature CPU-caches, the commonly cache incoherent \ac{dma} introduces a significant cost with the obligation to invalidate the transferred buffer cache lines.
In our case, the memory architecture requires the network driver to invalidate buffer cache lines prior to and after the processing by the \ac{nic} \ac{dma} engine, as shown in Figure \ref{fig:driver_buffer_flow}.
As cache management noticeably prolongs the execution time of Early Demultiplexing, we incorporate a lazy cache coherency establishment scheme into the driver to retain the highest possible performance as per \emph{Requirement 1.3}.
The driver is therefore split into two halves, as depicted in Figure \ref{fig:my_driver_buffer_flow_with_recycling}.

\begin{enumerate}[(1)]
    \item \textbf{Eager driver:} An immediately processed layer, executed as part of the \ac{isr}, classifies and enqueues packets.
    \item \textbf{Deferred driver:} A schedulable layer, executed in the network task according to the packet priority, establishes full cache coherency of received packets and prepares fresh replacement buffers.
\end{enumerate}

Prior to classification, only the first cache lines of the packet buffer containing the relevant header fields are invalidated.
Once the packet is chosen to be processed further, the remaining part is invalidated and a fresh packet buffer prepared and appended to the \ac{dma} \ac{bd} ring (cf. Section~\ref{sec:dma}).
This implies that as the differentiated flow queues fill up with packets, the \ac{bd} looses free packet buffers, forming a closed pool of packets shared by the \ac{bd} ring and the differentiated flow queues.
To prevent starvation of the \ac{bd}-ring caused by \ac{lp}-packets in the differentiated flow queues, the eager driver recycles lowest priority packet buffers once the \ac{bd}-ring hits a critical threshold (e.g. $\frac{1}{2}$).
This can be carried out with little cost, since only the accessed header cache lines have to be invalidated again.

\begin{figure}[h]
    \centering
    \includegraphics[scale=0.6]{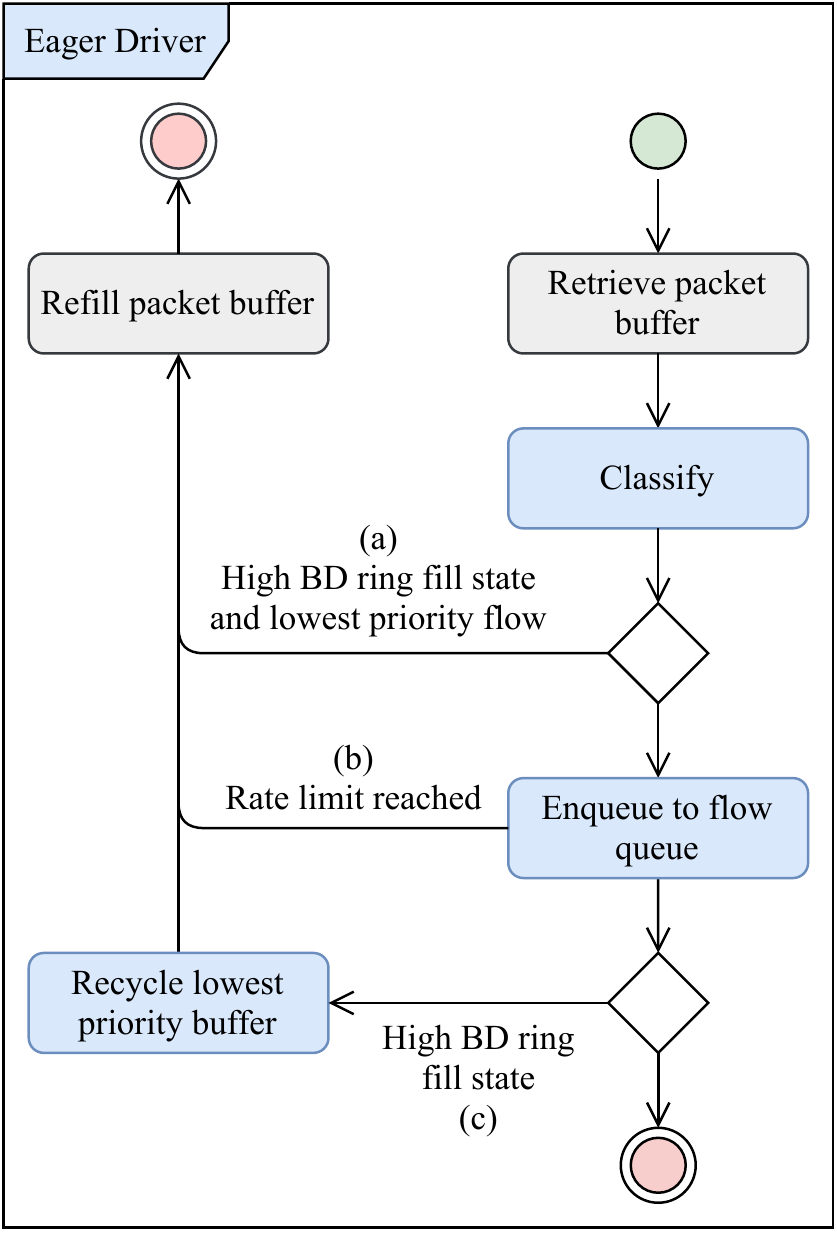}
    \caption{\textbf{Eager driver ISR:} Key execution paths that determine whether and when a packet buffer is recycled to save execution time in high-load scenarios.}
    \label{fig:top_half_activity}
\end{figure}

The resulting activity in the eager half driver is depicted in Figure \ref{fig:top_half_activity}.
Notable are the three different execution paths that might be taken:
If, due to a high \ac{bd} ring fill state a packet buffer has to be recycled, and the currently considered packet is of lowest priority, it gets recycled in a short-circuiting branch \textsuperscript{(a)}.
A flow queue may decline further packets to prevent overload by this particular flow \textsuperscript{(b)}.
Lastly, if the short-circuit branch was not taken but the \ac{bd} ring fill state is high, another packet buffer has to be recycled and inserted into the \ac{bd} ring \textsuperscript{(c)}.

\subsection{Prioritized Protocol Handling}
\label{sec:prioritedprotocolhandling}

Once the heterogeneous incoming packets are demultiplexed into differentiated flow queues, the protocol processing can be carried out with the receiver priority, as proposed in \cite{druschel1996lazy,lee_interrupt_2010}.

We apply a priority inheritance scheme to the single protocol processing task \cite{mercer1991evaluation}.
It allows the task scheduler to preempt the packet processing at any point in time. Additionally, it keeps a low footprint in terms of task resources and can be integrated into embedded network stacks that commonly use a single network task.

To implement this scheme, the priority of the network task has to be moderated depending on the currently processed packet and waiting packets, in order to avoid priority inversion.
Consider $\mathcal{F}$ as in Definition \ref{def:1}, $P(f)$ the priority of flow $f$ and $p(f)$ a packet of $f$. Let further $\mathcal{W}$ be the set of currently waiting packets and $\mathcal{E}$ the set of packets in processing. The priority of the network task ($P_\text{IP-task}$) must be assigned as follows.

\[
P_\text{IP-task} = \max ( P(f): f \in \mathcal{F} \; \land \; \exists \:  p(f) \in \mathcal{W} \cup \mathcal{E} )
\]

This assignment implies the network task priority is recomputed every time a packet gets queued or a packet has been processed.
On packet reception, the priority needs to be elevated \textit{iff} the respective flow priority is higher than the current priority assigned to the network task.
On finished packet processing, the priority needs to be decreased \textit{iff} the priority of the highest priority packet waiting in the differentiated flow queues is lower than the current network task priority.
This operation is supported by the ability of the differentiated flow queue data structure to efficiently provide the highest enqueued priority (reconsider Figure~\ref{fig:pdest_depq}).

It may appear that by using priority inheritance carried out per packet, we put an overly high computational burden on the fixed-priority task scheduler.
Yet, among all possible designs that involve the task scheduler in packet scheduling decisions by correctly signaling the current priority demand at each time, this design has the lowest scheduler data structure manipulation overhead:
Another design could use multiple processing tasks with constant priority, to which packets are assigned according to their flow.
The unblocking operation triggered when the first packet of a particular priority is enqueued then adds at least the same overhead --- the task has to be moved into the priority-respective ready task list, and moved out once blocked again.

In order to also have the deferrable parts of the driver processing scheduled according to packet flows, the networking task dequeues a highest priority packet buffer from the differentiated flow queues and executes the second half of the driver before continuing with the regular processing procedure.

\subsection{Rate Limitation}
To take advantage of Early Demultiplexing while at the same time keeping the system protected from overload conditions, deterministic mitigation techniques~\cite{danicki2021detecting} are applied to all but the low priority best-effort flows.
Additionally, the unconditionally executed \ac{isr} that demultiplexes incoming packets could incur a high load even if the subsequent scheduling cuts off further processing.
Hence, an additional, global rate limitation needs to be present.

To apply the rate limitation, each flow is scheduled by a conceptual aperiodic events server with each incoming packet being modelled as an aperiodic request.
In our prototype we use the deferrable server scheme (cf. Section~\ref{sec:aperiodic}).
Beyond the server capacity, packets are discarded.
For the individual flow queues, this happens as part of the inserting operation (reconsider Figure~\ref{fig:top_half_activity}), in order to avoid a situation with a paused \ac{hp} flow queue full of packets blocking all other processing.

To enforce a global rate limitation, once the capacity has been reached in one period, the driver processing switches from \ac{isr}-based execution to a polling driver task, staying in this mode until the capacity is not immediately reached at the begin of a period anymore.
When not processing packet receive \acp{irq} issued by the \ac{nic}, the \ac{bd} ring is filled until eventually packets are discarded by the \ac{nic}.

\subsection{Policy Integration}

In order to control the scheduling properties \emph{capacity}, \emph{period} and \emph{priority} for a flow, we expose these to the user for each socket via the \texttt{setsockopt}-\ac{api}.
Special flows such as for the protocols ARP, ICMP and those that are managed by the network stack, such as DNS and DHCP, can be configured using C macro definitions.
Similarly, the scheduling properties for the global rate limitation can be defined.

\subsection{Limitations}

The ability to proceed with deferred packet processing after a phase of higher system load depends on the number of available packet buffers.
As these buffers have to be prepared for immediate \ac{dma} operation and therefore a constant amount is dedicated to the lower levels of processing, additional memory might be necessary.

IP fragmentation cannot be dealt with properly in our architecture.
To demultiplex fragmented packets, their reassembly had to be done in the \ac{isr}, jeopardizing its \ac{wcet}.
This design treats all packet fragments as belonging to a background priority flow.
Yet, IP fragmentation is discouraged, as it introduces robustness, reliability and security issues \cite{kent1987fragmentation, gilad2011fragmentation}.

\section{Evaluation of the Software Approach}
\label{sec:ipstack_eval}

In this section we present empirical results collected from our prototypical IP stack implementation and subsequently discuss the effectiveness of the approach.

\subsection{Test Setup}

The test setup contains the FreeRTOS operating system with a modified FreeRTOS+TCP stack running on a Xilinx Zynq-7000 processing system containing a dual-core ARM Cortex A9.
Networking is done through a Gigabit-class Ethernet interface controlled by a Marvell 88E1518 \ac{phy} controller. Notable features are \ac{dma} and TX/RX-checksum offloading.
Measurements are taken on a single core.

\begin{samepage}
Two methods for measuring the effect on system load under high packet loads were pursued:

\begin{enumerate}
	\item \emph{Passive:} A background worker carries out CPU intensive work and monitors its performance.
	\item \emph{Active:} The software is instrumented to indicate notable events, i.e. task switches, \acp{irq}, and packet processing.
\end{enumerate}
\end{samepage}

The former is suitable for precisely estimating the average load that a particular scenario puts on the CPU.
While the latter introduces some overhead in the range of 1-5\% to the processing and misses some of the \ac{irq} switching, it allows us to evaluate the distribution of processing-induced latency.

\subsection{Experiment 1: CPU-Time Saved with Early Demultiplexing}
In this scenario two UDP sockets are bound, one with a low and one with a high priority receiver process.
To not alter the results, the capacity of all flows as well as the overall \ac{irq} limitation is set to infinity.

\begin{figure}[t]
    \centering
    \resizebox{0.7\columnwidth}{!}{
    \input{figures/cpu_time_per_packet_udp-zero.pgf}
    }
    \caption{\textbf{Processing impact: }CPU time per zero-length UDP packet under loads between $10^2$ and $10^6$ pkt/s with different configurations.\\
    ¹Short-circuiting branch disabled. ²Cache invalidation deferral disabled.}
    \label{fig:cpu_time_per_packet}
\end{figure}
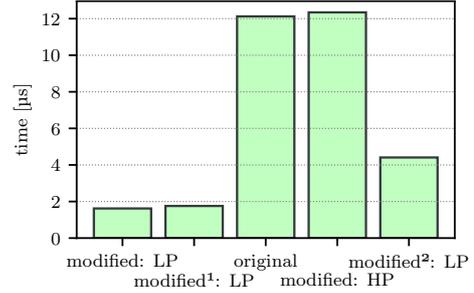

Multiple system configurations were confronted with a zero-length UDP-packet load of a constant rate for 60 seconds. Through passive measurement performed by a medium-priority task, the average CPU processing time per packet was then calculated (Figure \ref{fig:cpu_time_per_packet}).
In this experiment we observed that the CPU costs for processing a single packet are rather independent from the magnitude of incoming traffic, staying approximately constant in the range from $10^2$ to $10^6$ pkt/s.

\subsubsection*{Results}
The results show the difference in processing time between the packet processing paths. 
When \ac{lp} packets get no chance to be scheduled, the executed activity is only that of the Early Demultiplexing \ac{isr} with an average processing duration of $1.62 \mu s$ per packet.
Compared to the original stack as a baseline, which needs $12.1 \mu s$ to fully process a packet, this results in a speedup of 7.5x.
However, due to the short-circuiting logic depicted in Figure \ref{fig:top_half_activity} \textsuperscript{(a)}, in this constant \ac{lp}-flow measurement the packet buffers are discarded without being placed into a flow queue.
When disabling the short-circuiting code path, the per-packet processing time increases to $1.75 \mu s$, still yielding a seven-fold speedup compared to the full processing in the original stack.

In this scenario, the \ac{hp} packets are processed the entire network stack and cause a processing time of $12.3 \mu s$ per packet, decreasing receive performance by $1.7$ \% compared to the baseline stack.
This already small relative difference would decrease further if the subsequent (obligatory) reception by the receiver task was taken into account.

By modifying the prototype to again eagerly establish cache coherency in the \ac{isr}, the time spent for \ac{lp} packets increases notably to $4.4 \mu s$.
Hence, we conclude that incorporating a driver deferral mechanism into the architecture is essential to the performance on cached systems.

\subsection{Experiment 2: Packet Processing Latency}

The second experiment deals with the predictability of packet processing latencies in the modified IP stack.
Using the active approach, the reconstruction of precise execution times of each packet is possible.
Additionally, this allows us to differentiate between the execution paths of the modified driver.
Since we instrumented the \ac{isr} entry, some constant \ac{irq} overhead due to context saving is not included in this analysis.
Compared to Eperimant 1, where the overall impact on CPU time is measured, this experiment measures the duration of the eager driver per packet.

\begin{figure}[t]
    \centering
    \resizebox{0.7\columnwidth}{!}{
    \input{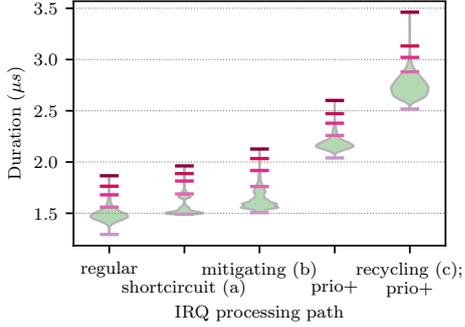}
    }
    \caption{\textbf{Eager driver: }Latency distributions for different execution paths in our modified stack:
    The horizontal bars indicate the percentiles $0\%$, $90\%$, $99\%$, $99.9\%$, $99.99\%$}
    \label{fig:packet_processing_latency}
\end{figure}

The system was flooded with $10^5$ zero-length UDP packets of two different priorities successively.
Figure~\ref{fig:packet_processing_latency} visualizes the distributions of \ac{isr} processing duration for specific processing paths. For each distribution, the quantiles $0\%, 90\%, 99\%, 99.9\%, 99.99\%$ are visualized as horizontal bars, in order to estimate a probabilistic \ac{wcet}.

\subsubsection*{Results}
\ac{lp} packets initially take the fastest path ("regular"), where incoming packets are enqueued without any other processing.
Once the \ac{bd} ring has reached a high fill state, packet buffers have to be recycled.
Since the incoming packets are already at the lowest level present in the differentiated flow queues, the short-circuiting path ("shortcircuit", \textsuperscript{(a)} in Figure \ref{fig:top_half_activity}) is taken.

\ac{hp} packets in contrast can cause a noticeable increase in \ac{isr} processing time.
At each occurrence of such a packet, the network task priority has to be increased in order to be scheduled subsequently ("prio+").
In case the \ac{bd} ring is already filled by previously received \ac{lp} packets now waiting inside their flow queue, a revocation is needed, adding further processing time ("recycling; prio+").
We also investigated on \ac{hp} packets that get rejected from their flow queue ("mitigating"), yet they behave similarly as shortcircuited packets.

The results show that the first three execution paths are similarly fast, while the ones that include an increase in priority or recycling operations are more costly.
As we discussed in section \ref{sec:prioritedprotocolhandling}, a priority increase can only happen if the flow priority of a received packet is higher than the one of all the currently enqueued ones.
Without the network task being active to process packets and lower the highest enqueued priority again, this is only possible once for each flow in a cascade of increasingly prioritized flows.
Thus, when the system is flooded for some time and \ac{lp} packets start building up in their queues, eventually the faster paths of the \ac{isr} will be taken.

\subsection{Experiment 3: Mitigation and Prioritization}

The final experiments show the effect of protocol processing prioritization and rate limitation, applied both for an individual flow and globally.
Experiments were conducted for multiple combinations of packet flood rates for a \ac{hp}- and \ac{lp}-flow, respectively, over a duration of 3 seconds each.
Again, a medium prioritized task measured the CPU load passively, preventing the scheduling of \ac{lp} packets.
Additionally, a receiver task was employed for the \ac{hp} flow in order to count the packets that arrived at their destination.
Figure \ref{fig:mitigation_map} shows the CPU utilization and the ratio of successfully received \ac{hp} packets to sent ones, as a function of both packet rates.

The original stack was slightly modified to feature an overall \ac{isr} rate limitation, in order to allow a meaningful comparison to our approach.
It is implemented by switching to polling mode once the capacity is reached for a certain period, similar to the one employed in our prototype. In this experiment, the limitations is set at 3 packets per 2 milliseconds.

\subsubsection*{Results}
The CPU utilization increases linearly along with both packet rates, until the global limit of 1500 pkt/s is reached.
Once the polling mode is active, the CPU load drops noticeably.
This can be accounted to the performance improvements gained by switching to a polling-based retrieving activity that handles multiple packets at once.
Further increasing the packet rate causes more \ac{hp} packets to be discarded by the \ac{nic}.

For the modified stack, parameter values anticipating a similar worst case CPU utilization were chosen.
We configured a high priority flow to allow one packet per millisecond and an unbounded low priority flow.
The \ac{isr} was limited to processing 7000 packets per second.

The CPU load also increases linearly with both packet rates.
As we would expect from the results of the first experiment, the load increases much slower with increasing \ac{lp} packet rates (notice the denser scale).
Above 1000 pkt/s (blue line) of \ac{hp} packets, the utilization stagnates as processing of further packets is cut off by the flow queue.
The additional triggered \ac{isr} executions are negligible at this scale.
When the sum of both rates exceeds 7000 pkt/s (black line), the CPU utilization also drops with polling activated.
Regarding the liveness of the \ac{hp} flow, we can see how it continuously decreases above the flow-specific rate of 1000 pkt/s.
Additionally, the global limitation impacts the \ac{hp} flow.
So, independent of the \ac{hp} flow rate itself, the communication liveness drops as the system is flooded with \ac{lp} packets.

\begin{figure}[h]
    \centering
    \resizebox{\columnwidth}{!}{
    \input{figures/mitigation_map.pgf}
    }
    \caption{\textbf{Results: }CPU utilization and \ac{hp} flow liveness at various packet rates on our modified system versus the original system employing only an overall rate limitation. \\
    The blue and black lines mark flow-specific and overall rate limitations respectively.}
    \label{fig:mitigation_map}
\end{figure}

When comparing the approach to a simple mitigating stack as a baseline, it becomes clear that it cannot help with processing higher rates of important packets.
Instead, supported by fast Early Demultiplexing and individual prioritization against the remaining tasks, it allows to postpone an overall limitation.
This way, a system can sustain a much higher load of less important packets before real-time disturbing effects start to occur.

\section{The Hardware Approach: Priority-Aware Interrupt Moderation}
\label{sec:nic}
Improving the prioritized real-time performance of network drivers still leaves a door open to overwhelm the system with network-generated interrupts. Since the \acp{irq} are triggered outside the operating system's sphere of influence, smartly moderating these interrupts requires some modification to the generating hardware, in this case the \ac{nic}. This section proposes an extension to the receive functionality of \acp{nic} to minimize \acp{irq} under high load while maintaining short packet receive delays for critical tasks. %

\begin{figure*}
\centering
\includegraphics[width=0.65\textwidth]{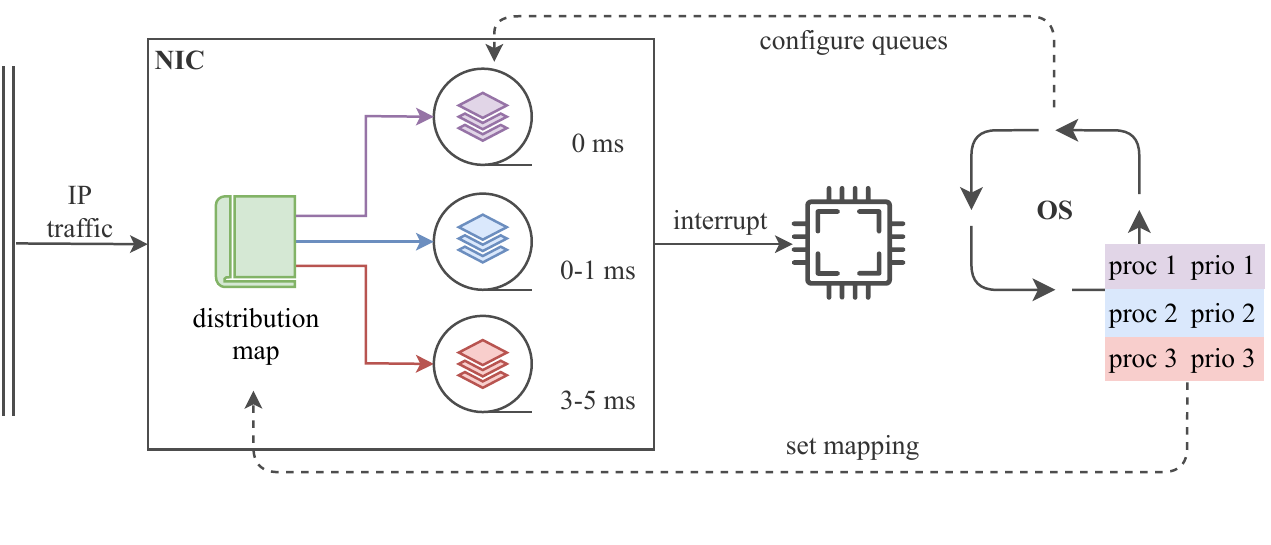}
\caption{\textbf{Multiqueue NIC:} Traffic is organized into different queues with exemplary delay values attached; configurations of queues and mapping are performed by the operating system upon socket binding.} \label{fig:design}
\end{figure*}

\subsection{Requirements}
\label{sec:requirements}
We specify four requirements to the \ac{nic} adaptation appropriate for real-time IoT devices with control over physical processes. 

\begin{enumerate}
\item[2.1] \emph{Interrupt Moderation.} The danger of network-generated interrupt floods and unpredictable networking overheads should be mitigated by reducing the number of interrupts triggered by the \ac{nic}. %
\item[2.2] \emph{Packet Prioritization.} Solely limiting the number of interrupts leads to an analogous increase in receive delays as packets are accumulated before a notification occurs. %
Furthermore, the system can still be flooded with unrelated packets forcing the operating system to handle them. 
Packets need to be classified and filtered before they reach the operating system.
\item[2.3] \emph{NIC Parametrization.} To effectively prioritize and filter packets in different environments, the introduced \ac{nic} needs to be configurable. It needs to be possible to affect generated receive delays per process and tune them for the scenario's real-time requirements.
\item[2.4] \emph{Continuous Configurability.} The necessary configuration of the \ac{nic} needs to be possible dynamically during runtime to facilitate changes in processes and the environment.
\end{enumerate}

\subsection{Overview}
The problem of network-generated interrupts affecting system performance can be solved by interrupt moderation (\emph{Requirement 2.1}). However, the techniques presented also have drawbacks. While they increase the overall efficiency of interrupt processing, they also increase the resulting packet delays and make them less predictable since packets are held back for a variable amount of time. In real-time systems, where process runtimes depend on incoming traffic, the occurrence of inaccuracies must be minimized. Therefore, a potential \ac{nic} design must attempt to reduce network overhead while guaranteeing low and constant latency for critical packets. As presented in Section~\ref{sec:related_nic} some specialized hardware exists running on FPGAs for specific real-time application types as well as multiqueue NICs for better multicore performance in data centers. However, to the best of our knowledge none exist for embedded IoT hardware or real-time processing in general IP networks. 

To this end, the interrupt moderation parameters are designed to be reconfigurable. By minimizing the relative packet delay, the time a packet dwells in memory before it is processed is reduced. At the same time, however, this increases the total number of interrupts, which reduces efficiency. This problem is not unique to real-time IoT devices, but is a natural consequence of interrupts. 
In closed IIoT environments, however, we can take advantage of this fact. Since embedded systems typically perform a fixed set of specific tasks, we can use their metadata to filter and manage incoming packets before interruption at the hardware level. These are the priorities of the protocols or packet-receiving processes and their associated IP flows. Interrupt moderation can thus become a tool to enforce priority-compliant traffic scheduling before they enter the realm of the operating system.

\subsection{Hardware Modifications}
\label{sec:adaptation}
The modifications made to the NIC concern only the reception of packets and begin after frame validation at the MAC layer. An illustration of the design can be seen in Figure \ref{fig:design}. To accommodate incoming packets belonging to different real-time processes, the receive buffer holding packet descriptors of the NIC is divided into multiple queues realized as ring buffers. This way, packet descriptors are assigned to different queues depending on their destination process and its priority, accounting for \emph{Requirement 2.2}. 

The metadata of validated packets is compared to a list of registered ports residing in a distribution map on the NIC. Here, packets are assigned to queues which hold packets of one IP flow each. According to the process priority and expected packet load, different interrupt moderation configurations (e.g. delay timers and counter threshold) are applied to them by the operating system. 

This way, packets for critical processes trigger interrupts immediately upon reception while less important packets (packets with low priority receiving tasks) are coalesced before one interrupt is triggered for all packets in the respective queue, indicated by the millisecond specifications in Figure \ref{fig:design}. Packets with no associated process can be dropped before an interrupt is triggered since these packets would be dropped by the operating system at a later point in any case, but after generating unnecessary ISR and network stack work. This is especially important under high unanticipated traffic loads targeting the device and potentially leading to a denial of service. Finding appropriate queue configurations for different process priorities is part of the design process of the embedded real-time system as introduced delays need to be part of scenario modeling.

\subsubsection{Synthetically Added Bursts}
\label{sec:bursts}
Coalescing packets in the NIC reduces the number of interrupts triggered, ISRs run and context switches performed. However, the amount of data to be processed by the network stack remains unchanged for packets registered for one of the running processes. Depending on the number of packets coalesced, interrupt moderation might lead to an accumulation of network stack workload into bursts. The necessary runtime to process an incoming packet is a lot smaller than the delay introduced by coalescing packets. A meaningful delay through these bursts can hence only happen under extremely high packet rates. We consider this when choosing the queue parameters as follows.

\subsubsection{Relevant Parameters}
\label{sec:parameters}
The multiqueue NIC introduces four parameters affecting packet delays and resource utilization as posed in \emph{Requirement 2.3}:

\begin{itemize}
\item \emph{Number of queues $m$.} The number of queues the receive buffer is divided into depends on the number of currently active processes accepting packets and supported protocols.
\item \emph{Size of a queue $n_q$.} The number of elements of a queue $q$ corresponds to its expected packet load, available memory, and moderation parameters.
\item \emph{Absolute queue timer values $t_\text{abs}(q)$.} Periodic duration until an interrupt is triggered by the queue $q$.
\item \emph{Packet timer values $t_\text{pack}(q)$.} Amount of time after a packet is received by the queue $q$ that triggers an interrupt if not reset by another incoming packet.
\end{itemize}

Additionally, the system introduces one implicite parameter:

\begin{itemize}
\item \emph{Maximum expected packet rate $R_\text{max}(q)$.} The maximum expected packet rate of the flow corresponding to a queue $q$. Equal to $\frac{1}{t_P(f)}$ as per Definition \ref{def:1}.
\end{itemize}

The timer values are used to span a time window of how long a packet remains in the queue. Depending on the packet rate, a variable number of packets is then coalesced to be announced by one interrupt. 
As these parameters have a high impact on the timeliness of incoming traffic and generated workload on the real-time device, the accuracy of their configuration is of high importance. While the number of queues is directly dependent on the current number of active (i.e. socket binding) processes, timer values and queue sizes have to be cautiously chosen. To be able to sensibly choose the parameters knowledge about expected packet rates and slack times of real-time processes is necessary.

The added packet processing time needs to be accounted for when developing an IP-connected real-time system. Process deadlines must allow for enough slack time for the system to handle concurrent packet reception. The higher the slack times are, the more packets can be processed without resulting in deadline misses. The worst-case scenario is subject to high interrupt rates affecting the process with the smallest slack time. For the calculation of appropriate queue parameters this value has to be factored in. The parameters must be chosen respecting the following considerations. 

\paragraph*{Queue size}
Choosing an appropriate queue size affects memory consumption as well as the maximum number of packets that can be coalesced in one interrupt. Applying interrupt moderation generally means holding more unprocessed packets in memory ultimately increasing the demand for the whole system. Memory implications for queue structures behave analogous but in a much smaller scale as only descriptors are held. 

The more packets can be held by one queue, the bigger the packet burst to be processed by the IP stack may become. Hence, this parameter also enforces an upper limit for the incoming packet rate per queue as elements are dropped when new packets arrive at a full queue. Additionally, this value has to be kept small enough for the IP stack processing time to be smaller than the minimum slack time when all queues generate a burst at the same time.   
The resulting delay corresponds to the Worst-Case Packet Processing Delay (WCPD) which depends on the per packet processing time $t_\text{netstack}$.

\[\text{WCPD} = t_\text{netstack} \sum_{q=0}^{m-1} n_q \]

\paragraph*{Absolute timer value}
The absolute queue timer realizes the upper latency bound of the interrupt rate window. To minimize the number of interrupts, this parameter needs to be maximized. At the same time, a higher absolute timer value also leads to a higher added latency an incoming packet might experience. Hence, the absolute timer value is limited by the maximum additional delay the underlying process can handle while still meeting its deadline. As a high value increases the burst of packets to be processed under high loads, the chance of the queue filling up increases, leading to packet loss. The maximum expected packet rate $R_\text{max}$ per queue $q$ needs to be factored in.
$t_d$ is the process-specific maximum allowed delay.

\[ t_\text{abs}(q) \le \max(t_{d}(q), \; t_{\text{qf}}(q)) - \text{WCPD}  \]
\[ t_{\text{qf}}(q) = \frac{n_q}{R_\text{max}(q)} \]

\paragraph*{Packet timer value}
The packet timer realizes the lower latency bound of the interrupt rate window. Increasing this value generally decreases the number of interrupts as there is more time available for a new packet to arrive and reset the timer. This also means, that this value directly influences the minimum additional latency an incoming packet experiences. At the same time the amount of interrupts is highly dependent on the incoming traffic shape. The worst case in terms of interrupts generated is a packet rate corresponding to the packet timer value (one interrupt per packet). Hence, the available slack time needs to suffice to handle the number of interrupts generated by all packet timers combined when every timer iteration of each queue triggers an interrupt. $t_P(f)$ is the expected packet arrival period as per Definition 1.

\[
t_P(f) \leq t_\text{pack}(q) \leq t_\text{abs}(q)
 \]

\subsubsection{Configuration}
As shown in Figure \ref{fig:design}, two interfaces are used for the NIC configurations. One for setting the queuing parameters mentioned earlier and a second to write process-to-IP flow mappings to the distribution map. Both configurations are performed when a socket is bound using the network stack API (see Section \ref{sec:rx_path}). For this purpose, the socket API is extended with driver calls that make the specific changes. Whenever a new process registers or releases a socket, the operating system transparently adjusts the number of queues and their parameters. The delay times and the size of the queues must be set for specific scenarios. The required information is passed to the driver as socket binding parameters.

As stated in Requirement 2.4, the system must be dynamically tunable at runtime to facilitate changes in processes or IP flows. Since the configuration process is linked to the socket API, the required tuning parameters can be passed at any time by the registering process. In the same way, NIC queues are released when a socket is no longer bound.

Since a network packet does not contain explicit information about the receiving process, a mapping must be made between the packet metadata and the processes. For this purpose, a mapping between IP flows and processes is created and placed on the NIC. In this design, the destination port is used to map a packet to a process. 
However, it could also be extended to match specific tags. This is be useful when all components of the distributed system are under the control of the developer to add security measures. The map must be on the NIC itself to cause minimal additional distribution delay, and still be configurable by the operating system to reflect the current set of existing processes.

\subsubsection{Memory Implications}
The presented approach has implications for memory usage on two levels: Firstly, the network buffer on the host system needs to be able to hold packet contents until they are processed, even when multiple packets are coalesced. Depending on the timer values and packet rate this might be a multiple of the usually necessary space. The network buffer resides on system RAM and is accessed by \ac{dma}. 
The second level is the required memory on the \ac{nic}. In an example implementation with 32~bit addresses, 32~bit timers, and a conservatively chosen maximum packet queue length of $65{,}536$ KB the on-NIC memory necessary for one table entry is 30 Bytes as broken down in Table~\ref{tab:memory}.

\begin{table}[h]
\centering
\caption{Required register memory per queue on the NIC for an example implementation.}
\begin{tabular}{l|c}
Component & Size \\ \hline
\emph{port\_id} & 16 bit \\ 
\emph{base\_address} & 32 bit \\
\emph{buffer\_size} & 16 bit \\
\emph{offset} & 16 bit  \\
\emph{next\_base\_address} & 32 bit \\
\emph{packet\_timer} & 32 bit \\
\emph{absolute\_timer} & 32 bit \\
\emph{packet\_timer\_expiration} & 32 bit \\
\emph{queue\_timer\_expiration} & 32 bit \\
\end{tabular}
\label{tab:memory}
\end{table}

The \emph{base\_address} field contains the RAM address of the beginning the the queue packet memory. The \emph{offset} is incremented for each incoming packet by its size. The \emph{next\_base\_address} is switched for the \emph{base\_address} when an interrupt for the queue occurs to be able to receive new packets while the buffered ones are processed.
When reserving one queue for non-transport layer protocols such as ARP (not requiring interrupt moderation fields and hence being 14 Bytes wide), the total memory requirement is

$$m \cdot 30 \text{ B} + 14 \text{ B}$$

for $m$ table entries on the NIC.

\section{Evaluation of the Hardware Approach}
\label{sec:nic_eval}
The proposed NIC extension reduces the number and frequency of interrupts caused by incoming packets. Yet, as packets that belong to registered processes are not dropped, driver and network stack workloads remain in a time shifted manner. 
We evaluate the resulting timing implications by conducting three sets of experiments: The first explores the ability to reduce interrupts. The second compares the robustness of the real-time system against high traffic loads. The third analyzes the effects of different queue configurations under expected loads.

\subsection{Test Setup}
\begin{figure}
\centering
\includegraphics[scale=0.9]{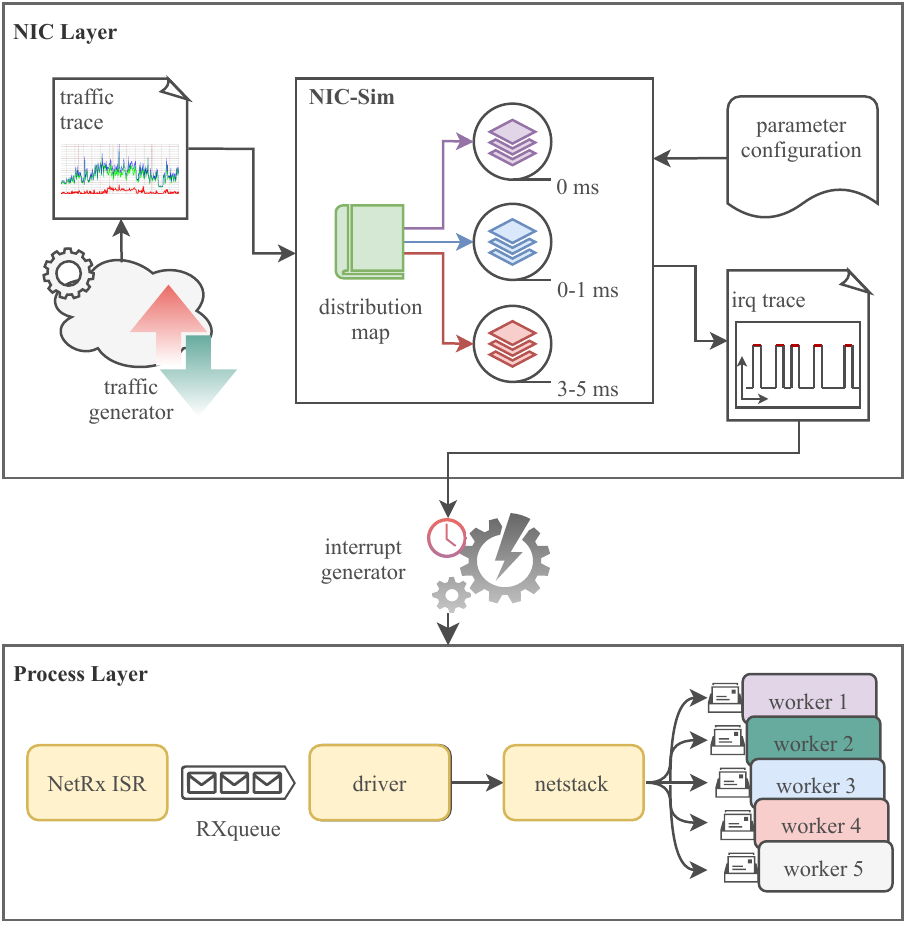}
\caption{\textbf{Evaluation setup on two layers:} Interrupt/packet trace generation by a NIC simulator and process observation on a real-time IoT device.} \label{fig:setup}
\end{figure}

As the design proposes changes to hardware but an evaluation on a real IoT device is necessary for plausibility, the evaluation platform comprises of two layers. One layer assuring plausibility by providing a real IoT device running real-time processes (process layer) and one allowing hardware and configuration changes to the NIC (NIC layer). The evaluation setup is depicted in Figure \ref{fig:setup}.

To evaluate the real-time behavior of a running IoT system, we used an ESP32\footnote{\url{https://www.espressif.com/en/products/socs/esp32}} microcontroller for the process layer. It is equipped with a dual-core CPU and widely used for IoT tasks that involve communication via WiFi and Bluetooth. The two cores of the device permit a separation between observed processes and testing system. The observed processes run on the real-time operating system FreeRTOS where scheduling is performed preemptively on basis of process priorities. To generate realistic network loads, traffic traces from common industrial control systems are used. 

\subsection{NIC Layer: Simulator Implementation}
The upper half of Figure \ref{fig:setup} illustrates the NIC layer. A traffic generator pre-processes captured network traces and synthetic load patterns to generate a receive traffic trace for the NIC simulator. The simulator is written in Python using the event-based simulation library \emph{SimPy}. 
It is configured for each experiment run (as explained in Section \ref{sec:adaptation}) depending on relevant IP flows and processes. 
With the possibility to change NIC parameters, different interrupt traces can be created from the same network packet stream.
These traces additionally contain packet metadata for use by the process layer.

\subsection{Process Layer: Network Stack Implementation}
The interrupt traces generated by the simulator are applied to the processing layer by an interrupt generator implemented on the ESP32.
Due to the interrupt moderation, each interrupt notifies the network driver of a batch of one or more incoming packets. The NIC interrupt service routine (NetRx ISR in Figure \ref{fig:setup}) receives this batch of packet descriptors and appends them to an operating system queue to be fetched by the network driver. From here, each packet is processed by the network stack task. If the packet destination port has a socket registered to it, the packet descriptor is forwarded to the socket mailbox and the associated task is notified. Otherwise, the packet is dropped.

The receiving real-time worker processes get access to sockets through the socket API. 
Using a receive function, the processes can then read data from the socket mailbox.
This approach implementation of Berkeley sockets corresponds to the mapping of IP flows to processes.
The receiving processes are workers of different priority. 
Each incoming packet is tracked from its time of arrival at the NIC until processing in its worker process where it triggers the task workload.

\subsection{Experiment 1: Interrupt Generation}

\begin{figure*}
\centering
  \includegraphics[width=0.8\textwidth]{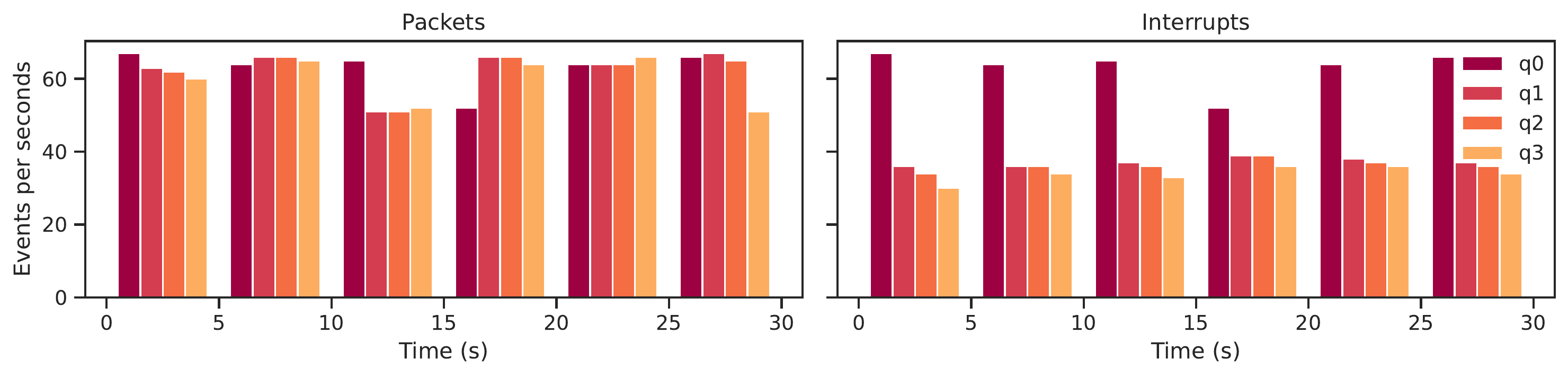}
  \caption{Histogram of packets and caused interrupts over time with bin size of 5 seconds.}
  \label{fig:pckirq}
\end{figure*}

In all experiments, the IoT device runs four worker processes of different priority. The processes receive traffic using the widely used industrial communication protocol MODBUS/TCP\footnote{MODBUS/TCP traces provided by \cite{frazao2018denial}.}. 
As each of the processes binds their own socket, four queues with different interrupt moderation configurations are set up in the NIC.

Table \ref{tab:configs} shows the queue configurations for the worker task flows. Queue 0 is configured to receive the IP flow of the critical task, hence no moderation is applied to this flow and packets are forwarded immediately upon arrival. Queues 1-3 are moderated with increasing delay values. Queue capacities are kept at a constant 128 packets per queue. Additionally, one baseline experiment is performed without any interrupt moderation. We observe the progression of interrupts generated in respect to packets received.

\begin{table}[h]
    \centering
    \caption{Queue configurations for baseline process IP flows in milliseconds.}
    \footnotesize
    \begin{tabular}{p{1cm}||p{1.5cm}|p{1.5cm}}
    	queue & absolute timer & packet timer\\
		\hline
        0 & \multicolumn{2}{c}{unmoderated} \\
		\cline{2-3}
		1 & 30 & 20 \\
        2 & 40 & 30 \\
        3 & 80 & 70 \\
    \end{tabular}
    \label{tab:configs}
\end{table}

\subsubsection*{Results}

The number of interrupts generated depends on the number of packets received in each queue and their configuration. 
Figure \ref{fig:pckirq} shows a comparison of packet and interrupt numbers for the baseline experiment without additional load. Queues 1 - 3 moderate interrupts in different time windows, so they generate fewer interrupts than queue 0, which is receiving packets for a critical task. 

\subsection{Experiment 2: Unfiltered Packet Flood}

To observe the system under high traffic, it is subjected to packet floods ranging from 0 to 15000 packets per second. The worker setup on the device stays as defined in Experiment 1. To observe the effects of packet floods when they are targeted at unregistered sockets they are subjected to a separate moderated NIC queue. We evaluate the compute load of the flood on the system.  The experiments are repeated on four different absolute delay timer values for the added packet floods.
The absolute timer values range from $800\,\mu s$ to $3200\,\mu s$ resulting in the designations \emph{nomod} (for unmoderated flood traffic), \emph{d800}, \emph{d1600}, \emph{d2400}, and \emph{d3200}. As we are testing the system under higher than expected load, the packet timer can be disregarded for this experiment (cf. Section \ref{sec:interrupt_moderation}). Each experiment runs for a duration of 30 seconds. 
We observe the additional runtime of the processes incurred by the network traffic.%

\subsubsection*{Results}

The total rate of interrupts per packet ranged from 70\,\% in the undisturbed experiment (Experiment 1) to 2\,\% with high additional load of 15000 packets per second and $3200\,\mu s$ absolute timer value. The absolute moderation timer is an effective tool to moderate the high load as more packets are coalesced into interrupts while the critical task is unaffected.

\begin{figure}[h]
\centering
\includegraphics[width=0.7\columnwidth]{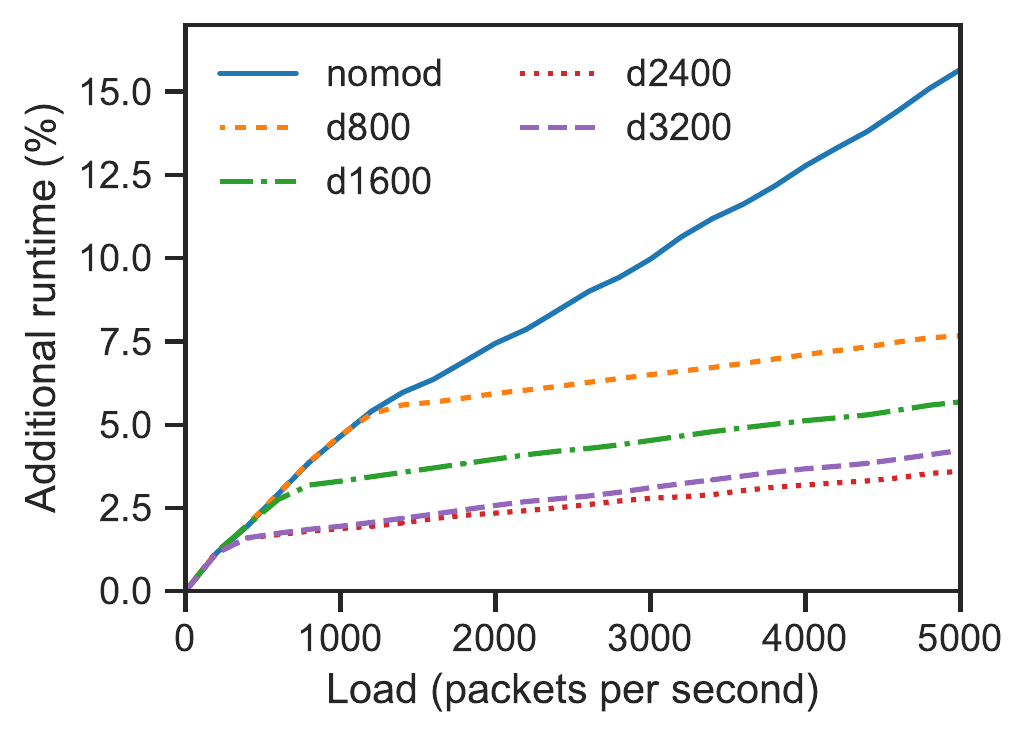}
\caption{Additional runtime of critical process (queue 0) induced by increased network load.\\\\\\}
\label{fig:runtimeq0}
\end{figure}

Next, we observe the interrupt-induced runtime increase of the critical process. A significant mitigation of the harmful effects of packet floods can be obtained in all moderation configurations for the critical process. Processes of lower priority also benefit from the approach, as the CPU is freed up from unnecessary ISRs. Figure \ref{fig:runtimeq0} shows the mitigating effects for the critical task under variable additional load. The visible linear increase continues throughout all experiments. Further, it can be seen that there is a scenario-specific optimal configuration between \emph{d2400} and \emph{d3200} due to the effects of packet burst processing (cf. Section~\ref{sec:bursts}). By increasing the delay parameters, more packets are coalesced for each interrupt, meaning that the networking tasks are confronted with larger bursts of packets per notification. This has negative effects on CPU load starting at a critical packet count. For the maximum depicted packet load of 5000 packets per second the additional runtime could be decreased by 80\,\% (or 12 percentage points) resulting from the prevention of 93\,\% of interrupts.

\subsection{Experiment 3: Expected Load}
Since the NIC retains packet notifications for low priority tasks, it causes an additional interrupt delay depending on the delay timer configurations. To investigate this delay, the second set of experiments was performed on a stable system with no unexpected traffic floods. The combination of absolute and packet delay timers spans a window for the period length of interrupts. Using the approximated incoming packet rate of a flow $1/t_P(f)$, developers can adjust the values to minimize the introduced interrupt delay as described in Section~\ref{sec:parameters}. 

In this experiment, the four processes receive IP flows ranging from about 50 (for queue 0) to 200 (for queue 3) packets per second. We compare three different NIC configurations:

\begin{itemize}
	\item \emph{No moderation}: All queues trigger interrupts as soon as a packet arrives.
	\item \emph{Medium moderation}: Queue 0 triggers an interrupt as soon as a packet arrives. Queues 1, 2, and 3 are configured to coalesce 2, 3, and 4 packets per interrupt, on average.
	\item \emph{Strict moderation}: Queue 0 triggers an interrupt as soon as a packet arrives. Queues 1, 2, and 3 are configured to coalesce 5, 6, and 7 packets per interrupt, on average.
\end{itemize}

\subsubsection*{Results}

\begin{figure}
\centering
\includegraphics[width=0.7\columnwidth]{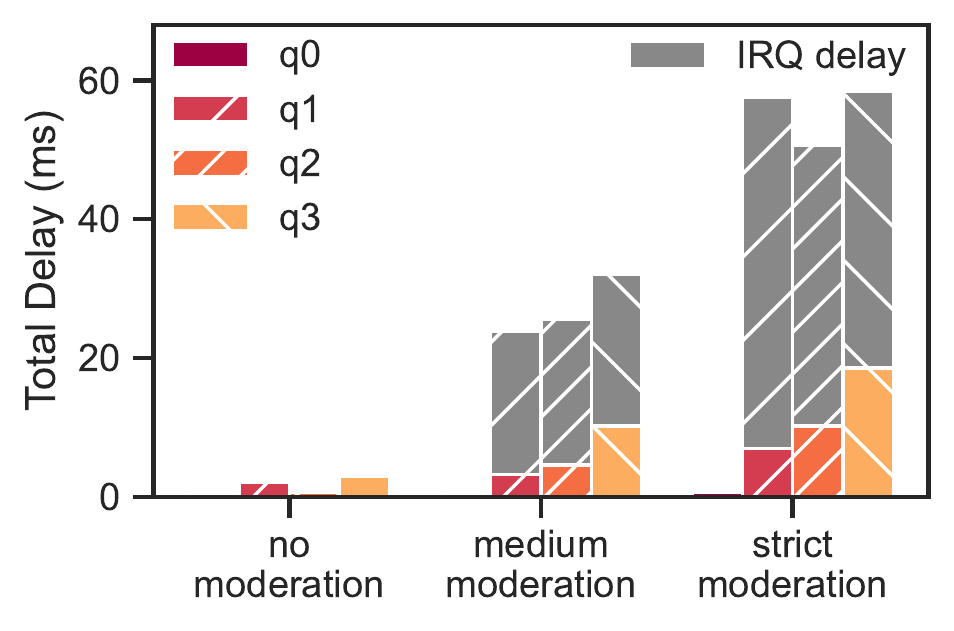}
\caption{Median delay times of the four configurations. The IRQ delay specifies the average coalescing delays.}
\label{fig:nodos}
\end{figure}

The results, depicted in Figure \ref{fig:nodos}, show that the critical process (queue 0) does not suffer from any additional delay.
The delay added to processes of less priority is directly dependent on the chosen moderation parameters and process-specific load and can be chosen to fit the requirements of each process before or during runtime.
At the same time, the delay window prevents interrupt frequencies to climb to a critical level.
An increase in the OS-induced delay can be observed when too many packets are coalesced to one interrupt as the resulting bursts in packet processing increase the processing time for individual packets in the networking and worker processes.

While the queue moderation can effectively reduce the impact of high packet loads, the tuning of moderation parameters requires care.
Task deadlines, as well as packet loads and latencies for each process should be identified to reduce the impact of interrupt moderation under normal conditions while at the same time ensuring operability for critical tasks under unexpectedly high loads.

\section{Towards a Unified Architecture for Real-Time Aware Packet Processing}
\label{sec:outlook}
In this section, we give an outlook on a combination of the two evaluated approaches, attending to challenges arising and necessary changes to be made to the individual designs. We furthermore discuss the compatibility of advantages gained through a unified hardware/software co-design for real-time packet reception.

\begin{figure*}[ht]
    \centering
    \includegraphics[width=0.85\textwidth]{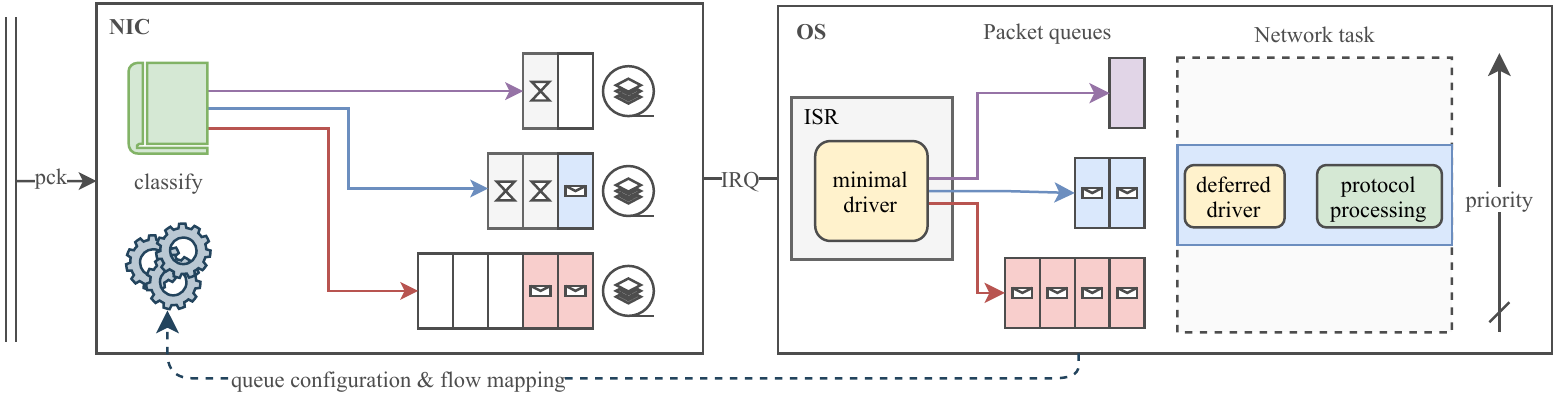}
    \caption{\textbf{Unified design:} Demultiplexing of packets and interrupt moderation in the NIC. Packet ordering, priority inheritance, and burst control in the network driver.}
    \label{fig:unified_design}
\end{figure*}

\subsection{Preliminary Considerations}
To implement a unified design, the networking task, driver, and hardware components should be designed holistically and make available configuration parameters in a standardized manner. This design is subject to the same assumptions and requirements as the individual approaches. Most notably the placement of devices in a real-time IoT setting with no industrial grade real-time networking infrastructure. The preliminary considerations in Section \ref{sec:considerations} remain relevant here.

A unified design could combine the advantages of the two presented approaches while also mitigate some of the individual disadvantages. These shortcomings can be summarized as follows. 

\paragraph{Hardware Approach}
\begin{itemize}
    \item Only the amount of ISR runs is reduced in a prioritized manner. Once packets are received by the OS queue, no reordering can happen and packet processing can still lead to an inversion of priorities.
    \item The design creates bursts of low priority packets that, once inside the operating system, might block the processing of high priority packets and lead to priority inversion.
    \item Interrupt moderation and multiqueue parameters have to be set by developers apriori and for each process individually.
\end{itemize}

\paragraph{Sofware Approach}
\begin{itemize}
    \item The number of interrupts can not be reduced other than by switching to polling mode. 
    \item In the case of a flood of incoming unwanted packets the system can still be overwhelmed by IRQs and ISR runs or be forced into polling mode with the remaining packet processing overhead.
    \item The classification of packets and packet-wise cache invalidation adds workload to the (preempting) ISRs.
\end{itemize}

\subsection{Overview}

The combined design unifies the priority spaces of real-time tasks, packet processing, and network-generated interrupts. An abstract representation of a unified design can be seen in Figure \ref{fig:unified_design}. Depending on developer-defined tasks and their priorities incoming packets can be dropped and coalesced to fewer IRQs before any workload emerges in the RTOS. With the mapping of tasks to IP flows being performed in the NIC, the packet classification is shifted from the eager part of the driver to hardware. The moderated queues in the NIC remain configurable over the socket API and write their contents to the \ac{bd} ring where the remaining part of the eager driver fetches them. The descriptors can then be placed in process specific packet queues enforcing the rate limitation and priority inheritance introduced by the software approach.   

\subsubsection{Changes to Individual Implementations}
To accommodate the combined design the individual implementations have to be adapted. This section discusses the challenges of a new design and presents necessary changes.

\paragraph{Hardware Approach}
Since the classification of packets should not have to be repeated in the driver, the packet descriptors written to the RTOS need to b extended to contain the priority of the expecting process. By furthermore adding a field for the IP-flow ID (i.e. the port number) flows of equal priority can be prevented from blocking each other in case one experiences a packet flood. This way, packet descriptors fetched from the ring buffer already contain priorities and flow IDs and only need to be enqueued accordingly by the ISR. 

A more complex yet optional change to the original design could include the dynamization of queue configuration. As a means of reactive rate limitation the driver could be enabled to dynamically change NIC queue sizes during periods of high packet volume. This way, packets could be dropped in a priority aware manner before reaching the driver when the rate limitation is expected to lead to the same. The initial configuration via the socket API can remain unchanged.

\begin{figure}[h]
    \centering
    \includegraphics[width=\columnwidth]{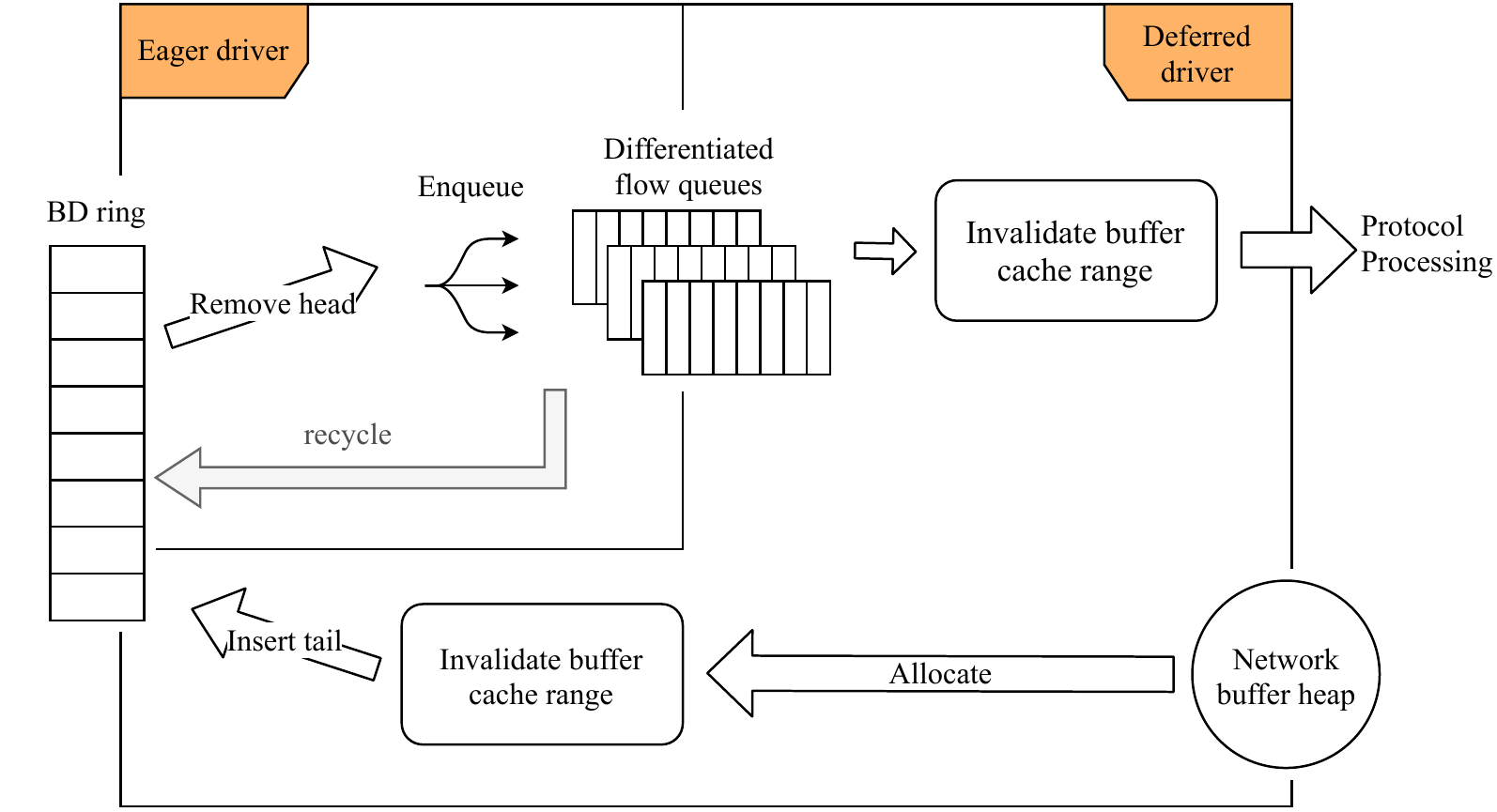}
    \caption{\textbf{Receive activity in adapted driver:} Classification, as well as cache invalidation of the header line is removed.}
    \label{fig:unified_driver}
\end{figure}

\paragraph{Software Approach}
With the classification being shifted to hardware, the software side can save packet processing time at the eager driver in two manners: The ISR does not have to read the packet headers from memory in order to place the descriptors in the appropriate queue as the descriptors now contain priority and flow ID fields. With header inspection in the ISR being unnecessary, the time consuming task of packet-wise cache line invalidation can also be eliminated from the eager driver, resulting in the receive workflow in Figure \ref{fig:unified_driver}. 

The due to the interrupt moderation bursty nature of packet reception through ISRs is not impeded by the combination of the two approaches. Packets, especially those of low priorities come in bursts, possibly resulting in immediate rate limitation by the driver. To prevent this from happening for each burst, queue sizes need to be adapted to each other.
Furthermore, some of the unschedulable ISR runs will require more time, as bursts of packets need to be handled. However, this is somewhat mitigated as the comparison and raise of the network task priority only has to happen once at the end of the ISR and not for every packet as in the original design.

\section{Conclusion}
\label{sec:conclusion}

Unexpected floods of network traffic can delay process flows in real-time systems, putting critical real-time requirements at risk. This paper presented two individually evaluated approaches -- one software- and one hardware-based -- to mitigate the real-time violating effects of IP packet reception in constrained IoT devices. 

The software approach is soft IP stack design to individually schedule packet processing for differently prioritized IP-flows after early demultiplexing.
The issue of costly processing in the network driver is approached by integrating the possibility of deferred buffer processing into our architecture.
On our test system, even when having to deal with packet buffers travelling CPU-caches, the CPU load caused by \ac{lp} packets in an already occupied system is reduced by $86\%$, leading to a 7-fold speedup of concurrently running processes.
Through limitation parameters, the software approach allows system designers to anticipate packet rates of certain soft real-time flows and derive an estimate for the respective processing \ac{wcet}.
Budgeting the same CPU resources to the processing of incoming packets, the networking subsystem can still process packets of a \ac{hp}-flow for up to $600\%$ higher overall traffic loads.

The hardware approach presented a multiqueue NIC design shifting the early demultiplexing to hardware. This way, the number of network-triggered IRQs can be moderated depending on packet priorities. By configuring the extended NIC via an adapted Berkeley socket API, currently active processes waiting for packets can register their priorities with the hardware, effectively mapping IP flows to processes when a socket is bound and changing queue parameters accordingly. 
We evaluated the hardware design using a NIC simulation and an IoT device running a real-time operating system. The results of these experiments show that our approach significantly reduces the impact of traffic floods on critical process runtimes by saving 93\,\% of interrupts and 80\,\% of processing delay under packet rates of 5000 per second while the configuration of multiqueue parameters requires knowledge about expected network traffic and real-time requirements.

The combination of the two approaches requires a hardware/software co-design of real-time NIC, network driver and IP stack implementation, further addressing the remaining weaknesses of the individual designs. 

\section*{Acknowledgments}
We sincerely thank the reviewers for their helpful comments which significantly improved this work. This research was supported by the German Academic Exchange Service (DAAD) as ide3a.

%% file: figures/cpu_time_per_packet_udp-zero.pgf
\begingroup%
\makeatletter%
\begin{pgfpicture}%
\pgfpathrectangle{\pgfpointorigin}{\pgfqpoint{2.760713in}{1.760000in}}%
\pgfusepath{use as bounding box, clip}%
\begin{pgfscope}%
\pgfsetbuttcap%
\pgfsetmiterjoin%
\definecolor{currentfill}{rgb}{1.000000,1.000000,1.000000}%
\pgfsetfillcolor{currentfill}%
\pgfsetlinewidth{0.000000pt}%
\definecolor{currentstroke}{rgb}{1.000000,1.000000,1.000000}%
\pgfsetstrokecolor{currentstroke}%
\pgfsetdash{}{0pt}%
\pgfpathmoveto{\pgfqpoint{0.000000in}{0.000000in}}%
\pgfpathlineto{\pgfqpoint{2.760713in}{0.000000in}}%
\pgfpathlineto{\pgfqpoint{2.760713in}{1.760000in}}%
\pgfpathlineto{\pgfqpoint{0.000000in}{1.760000in}}%
\pgfpathlineto{\pgfqpoint{0.000000in}{0.000000in}}%
\pgfpathclose%
\pgfusepath{fill}%
\end{pgfscope}%
\begin{pgfscope}%
\pgfsetbuttcap%
\pgfsetmiterjoin%
\definecolor{currentfill}{rgb}{1.000000,1.000000,1.000000}%
\pgfsetfillcolor{currentfill}%
\pgfsetlinewidth{0.000000pt}%
\definecolor{currentstroke}{rgb}{0.000000,0.000000,0.000000}%
\pgfsetstrokecolor{currentstroke}%
\pgfsetstrokeopacity{0.000000}%
\pgfsetdash{}{0pt}%
\pgfpathmoveto{\pgfqpoint{0.381946in}{0.314043in}}%
\pgfpathlineto{\pgfqpoint{2.670942in}{0.314043in}}%
\pgfpathlineto{\pgfqpoint{2.670942in}{1.760000in}}%
\pgfpathlineto{\pgfqpoint{0.381946in}{1.760000in}}%
\pgfpathlineto{\pgfqpoint{0.381946in}{0.314043in}}%
\pgfpathclose%
\pgfusepath{fill}%
\end{pgfscope}%
\begin{pgfscope}%
\pgfpathrectangle{\pgfqpoint{0.381946in}{0.314043in}}{\pgfqpoint{2.288996in}{1.445957in}}%
\pgfusepath{clip}%
\pgfsetbuttcap%
\pgfsetmiterjoin%
\definecolor{currentfill}{rgb}{0.000000,1.000000,0.000000}%
\pgfsetfillcolor{currentfill}%
\pgfsetfillopacity{0.250980}%
\pgfsetlinewidth{1.003750pt}%
\definecolor{currentstroke}{rgb}{0.211765,0.223529,0.239216}%
\pgfsetstrokecolor{currentstroke}%
\pgfsetdash{}{0pt}%
\pgfpathmoveto{\pgfqpoint{0.485991in}{0.314043in}}%
\pgfpathlineto{\pgfqpoint{0.832809in}{0.314043in}}%
\pgfpathlineto{\pgfqpoint{0.832809in}{0.494796in}}%
\pgfpathlineto{\pgfqpoint{0.485991in}{0.494796in}}%
\pgfpathlineto{\pgfqpoint{0.485991in}{0.314043in}}%
\pgfpathclose%
\pgfusepath{stroke,fill}%
\end{pgfscope}%
\begin{pgfscope}%
\pgfpathrectangle{\pgfqpoint{0.381946in}{0.314043in}}{\pgfqpoint{2.288996in}{1.445957in}}%
\pgfusepath{clip}%
\pgfsetbuttcap%
\pgfsetmiterjoin%
\definecolor{currentfill}{rgb}{0.000000,1.000000,0.000000}%
\pgfsetfillcolor{currentfill}%
\pgfsetfillopacity{0.250980}%
\pgfsetlinewidth{1.003750pt}%
\definecolor{currentstroke}{rgb}{0.211765,0.223529,0.239216}%
\pgfsetstrokecolor{currentstroke}%
\pgfsetdash{}{0pt}%
\pgfpathmoveto{\pgfqpoint{0.919513in}{0.314043in}}%
\pgfpathlineto{\pgfqpoint{1.266331in}{0.314043in}}%
\pgfpathlineto{\pgfqpoint{1.266331in}{0.509953in}}%
\pgfpathlineto{\pgfqpoint{0.919513in}{0.509953in}}%
\pgfpathlineto{\pgfqpoint{0.919513in}{0.314043in}}%
\pgfpathclose%
\pgfusepath{stroke,fill}%
\end{pgfscope}%
\begin{pgfscope}%
\pgfpathrectangle{\pgfqpoint{0.381946in}{0.314043in}}{\pgfqpoint{2.288996in}{1.445957in}}%
\pgfusepath{clip}%
\pgfsetbuttcap%
\pgfsetmiterjoin%
\definecolor{currentfill}{rgb}{0.000000,1.000000,0.000000}%
\pgfsetfillcolor{currentfill}%
\pgfsetfillopacity{0.250980}%
\pgfsetlinewidth{1.003750pt}%
\definecolor{currentstroke}{rgb}{0.211765,0.223529,0.239216}%
\pgfsetstrokecolor{currentstroke}%
\pgfsetdash{}{0pt}%
\pgfpathmoveto{\pgfqpoint{1.353035in}{0.314043in}}%
\pgfpathlineto{\pgfqpoint{1.699853in}{0.314043in}}%
\pgfpathlineto{\pgfqpoint{1.699853in}{1.666597in}}%
\pgfpathlineto{\pgfqpoint{1.353035in}{1.666597in}}%
\pgfpathlineto{\pgfqpoint{1.353035in}{0.314043in}}%
\pgfpathclose%
\pgfusepath{stroke,fill}%
\end{pgfscope}%
\begin{pgfscope}%
\pgfpathrectangle{\pgfqpoint{0.381946in}{0.314043in}}{\pgfqpoint{2.288996in}{1.445957in}}%
\pgfusepath{clip}%
\pgfsetbuttcap%
\pgfsetmiterjoin%
\definecolor{currentfill}{rgb}{0.000000,1.000000,0.000000}%
\pgfsetfillcolor{currentfill}%
\pgfsetfillopacity{0.250980}%
\pgfsetlinewidth{1.003750pt}%
\definecolor{currentstroke}{rgb}{0.211765,0.223529,0.239216}%
\pgfsetstrokecolor{currentstroke}%
\pgfsetdash{}{0pt}%
\pgfpathmoveto{\pgfqpoint{1.786557in}{0.314043in}}%
\pgfpathlineto{\pgfqpoint{2.133375in}{0.314043in}}%
\pgfpathlineto{\pgfqpoint{2.133375in}{1.691145in}}%
\pgfpathlineto{\pgfqpoint{1.786557in}{1.691145in}}%
\pgfpathlineto{\pgfqpoint{1.786557in}{0.314043in}}%
\pgfpathclose%
\pgfusepath{stroke,fill}%
\end{pgfscope}%
\begin{pgfscope}%
\pgfpathrectangle{\pgfqpoint{0.381946in}{0.314043in}}{\pgfqpoint{2.288996in}{1.445957in}}%
\pgfusepath{clip}%
\pgfsetbuttcap%
\pgfsetmiterjoin%
\definecolor{currentfill}{rgb}{0.000000,1.000000,0.000000}%
\pgfsetfillcolor{currentfill}%
\pgfsetfillopacity{0.250980}%
\pgfsetlinewidth{1.003750pt}%
\definecolor{currentstroke}{rgb}{0.211765,0.223529,0.239216}%
\pgfsetstrokecolor{currentstroke}%
\pgfsetdash{}{0pt}%
\pgfpathmoveto{\pgfqpoint{2.220079in}{0.314043in}}%
\pgfpathlineto{\pgfqpoint{2.566897in}{0.314043in}}%
\pgfpathlineto{\pgfqpoint{2.566897in}{0.805885in}}%
\pgfpathlineto{\pgfqpoint{2.220079in}{0.805885in}}%
\pgfpathlineto{\pgfqpoint{2.220079in}{0.314043in}}%
\pgfpathclose%
\pgfusepath{stroke,fill}%
\end{pgfscope}%
\begin{pgfscope}%
\pgfsetbuttcap%
\pgfsetroundjoin%
\definecolor{currentfill}{rgb}{0.000000,0.000000,0.000000}%
\pgfsetfillcolor{currentfill}%
\pgfsetlinewidth{0.803000pt}%
\definecolor{currentstroke}{rgb}{0.000000,0.000000,0.000000}%
\pgfsetstrokecolor{currentstroke}%
\pgfsetdash{}{0pt}%
\pgfsys@defobject{currentmarker}{\pgfqpoint{0.000000in}{-0.048611in}}{\pgfqpoint{0.000000in}{0.000000in}}{%
\pgfpathmoveto{\pgfqpoint{0.000000in}{0.000000in}}%
\pgfpathlineto{\pgfqpoint{0.000000in}{-0.048611in}}%
\pgfusepath{stroke,fill}%
}%
\begin{pgfscope}%
\pgfsys@transformshift{0.659400in}{0.314043in}%
\pgfsys@useobject{currentmarker}{}%
\end{pgfscope}%
\end{pgfscope}%
\begin{pgfscope}%
\definecolor{textcolor}{rgb}{0.000000,0.000000,0.000000}%
\pgfsetstrokecolor{textcolor}%
\pgfsetfillcolor{textcolor}%
\pgftext[x=0.659400in,y=0.216821in,,top]{\color{textcolor}\rmfamily\fontsize{8.000000}{9.600000}\selectfont modified: LP}%
\end{pgfscope}%
\begin{pgfscope}%
\pgfsetbuttcap%
\pgfsetroundjoin%
\definecolor{currentfill}{rgb}{0.000000,0.000000,0.000000}%
\pgfsetfillcolor{currentfill}%
\pgfsetlinewidth{0.803000pt}%
\definecolor{currentstroke}{rgb}{0.000000,0.000000,0.000000}%
\pgfsetstrokecolor{currentstroke}%
\pgfsetdash{}{0pt}%
\pgfsys@defobject{currentmarker}{\pgfqpoint{0.000000in}{-0.048611in}}{\pgfqpoint{0.000000in}{0.000000in}}{%
\pgfpathmoveto{\pgfqpoint{0.000000in}{0.000000in}}%
\pgfpathlineto{\pgfqpoint{0.000000in}{-0.048611in}}%
\pgfusepath{stroke,fill}%
}%
\begin{pgfscope}%
\pgfsys@transformshift{1.092922in}{0.314043in}%
\pgfsys@useobject{currentmarker}{}%
\end{pgfscope}%
\end{pgfscope}%
\begin{pgfscope}%
\definecolor{textcolor}{rgb}{0.000000,0.000000,0.000000}%
\pgfsetstrokecolor{textcolor}%
\pgfsetfillcolor{textcolor}%
\pgftext[x=1.092922in,y=0.098766in,,top]{\color{textcolor}\rmfamily\fontsize{8.000000}{9.600000}\selectfont modified¹: LP}%
\end{pgfscope}%
\begin{pgfscope}%
\pgfsetbuttcap%
\pgfsetroundjoin%
\definecolor{currentfill}{rgb}{0.000000,0.000000,0.000000}%
\pgfsetfillcolor{currentfill}%
\pgfsetlinewidth{0.803000pt}%
\definecolor{currentstroke}{rgb}{0.000000,0.000000,0.000000}%
\pgfsetstrokecolor{currentstroke}%
\pgfsetdash{}{0pt}%
\pgfsys@defobject{currentmarker}{\pgfqpoint{0.000000in}{-0.048611in}}{\pgfqpoint{0.000000in}{0.000000in}}{%
\pgfpathmoveto{\pgfqpoint{0.000000in}{0.000000in}}%
\pgfpathlineto{\pgfqpoint{0.000000in}{-0.048611in}}%
\pgfusepath{stroke,fill}%
}%
\begin{pgfscope}%
\pgfsys@transformshift{1.526444in}{0.314043in}%
\pgfsys@useobject{currentmarker}{}%
\end{pgfscope}%
\end{pgfscope}%
\begin{pgfscope}%
\definecolor{textcolor}{rgb}{0.000000,0.000000,0.000000}%
\pgfsetstrokecolor{textcolor}%
\pgfsetfillcolor{textcolor}%
\pgftext[x=1.526444in,y=0.216821in,,top]{\color{textcolor}\rmfamily\fontsize{8.000000}{9.600000}\selectfont original}%
\end{pgfscope}%
\begin{pgfscope}%
\pgfsetbuttcap%
\pgfsetroundjoin%
\definecolor{currentfill}{rgb}{0.000000,0.000000,0.000000}%
\pgfsetfillcolor{currentfill}%
\pgfsetlinewidth{0.803000pt}%
\definecolor{currentstroke}{rgb}{0.000000,0.000000,0.000000}%
\pgfsetstrokecolor{currentstroke}%
\pgfsetdash{}{0pt}%
\pgfsys@defobject{currentmarker}{\pgfqpoint{0.000000in}{-0.048611in}}{\pgfqpoint{0.000000in}{0.000000in}}{%
\pgfpathmoveto{\pgfqpoint{0.000000in}{0.000000in}}%
\pgfpathlineto{\pgfqpoint{0.000000in}{-0.048611in}}%
\pgfusepath{stroke,fill}%
}%
\begin{pgfscope}%
\pgfsys@transformshift{1.959966in}{0.314043in}%
\pgfsys@useobject{currentmarker}{}%
\end{pgfscope}%
\end{pgfscope}%
\begin{pgfscope}%
\definecolor{textcolor}{rgb}{0.000000,0.000000,0.000000}%
\pgfsetstrokecolor{textcolor}%
\pgfsetfillcolor{textcolor}%
\pgftext[x=1.959966in,y=0.098766in,,top]{\color{textcolor}\rmfamily\fontsize{8.000000}{9.600000}\selectfont modified: HP}%
\end{pgfscope}%
\begin{pgfscope}%
\pgfsetbuttcap%
\pgfsetroundjoin%
\definecolor{currentfill}{rgb}{0.000000,0.000000,0.000000}%
\pgfsetfillcolor{currentfill}%
\pgfsetlinewidth{0.803000pt}%
\definecolor{currentstroke}{rgb}{0.000000,0.000000,0.000000}%
\pgfsetstrokecolor{currentstroke}%
\pgfsetdash{}{0pt}%
\pgfsys@defobject{currentmarker}{\pgfqpoint{0.000000in}{-0.048611in}}{\pgfqpoint{0.000000in}{0.000000in}}{%
\pgfpathmoveto{\pgfqpoint{0.000000in}{0.000000in}}%
\pgfpathlineto{\pgfqpoint{0.000000in}{-0.048611in}}%
\pgfusepath{stroke,fill}%
}%
\begin{pgfscope}%
\pgfsys@transformshift{2.393488in}{0.314043in}%
\pgfsys@useobject{currentmarker}{}%
\end{pgfscope}%
\end{pgfscope}%
\begin{pgfscope}%
\definecolor{textcolor}{rgb}{0.000000,0.000000,0.000000}%
\pgfsetstrokecolor{textcolor}%
\pgfsetfillcolor{textcolor}%
\pgftext[x=2.393488in,y=0.216821in,,top]{\color{textcolor}\rmfamily\fontsize{8.000000}{9.600000}\selectfont modified²: LP}%
\end{pgfscope}%
\begin{pgfscope}%
\pgfpathrectangle{\pgfqpoint{0.381946in}{0.314043in}}{\pgfqpoint{2.288996in}{1.445957in}}%
\pgfusepath{clip}%
\pgfsetbuttcap%
\pgfsetroundjoin%
\pgfsetlinewidth{0.501875pt}%
\definecolor{currentstroke}{rgb}{0.501961,0.501961,0.501961}%
\pgfsetstrokecolor{currentstroke}%
\pgfsetdash{{0.500000pt}{0.825000pt}}{0.000000pt}%
\pgfpathmoveto{\pgfqpoint{0.381946in}{0.314043in}}%
\pgfpathlineto{\pgfqpoint{2.670942in}{0.314043in}}%
\pgfusepath{stroke}%
\end{pgfscope}%
\begin{pgfscope}%
\pgfsetbuttcap%
\pgfsetroundjoin%
\definecolor{currentfill}{rgb}{0.000000,0.000000,0.000000}%
\pgfsetfillcolor{currentfill}%
\pgfsetlinewidth{0.803000pt}%
\definecolor{currentstroke}{rgb}{0.000000,0.000000,0.000000}%
\pgfsetstrokecolor{currentstroke}%
\pgfsetdash{}{0pt}%
\pgfsys@defobject{currentmarker}{\pgfqpoint{-0.048611in}{0.000000in}}{\pgfqpoint{-0.000000in}{0.000000in}}{%
\pgfpathmoveto{\pgfqpoint{-0.000000in}{0.000000in}}%
\pgfpathlineto{\pgfqpoint{-0.048611in}{0.000000in}}%
\pgfusepath{stroke,fill}%
}%
\begin{pgfscope}%
\pgfsys@transformshift{0.381946in}{0.314043in}%
\pgfsys@useobject{currentmarker}{}%
\end{pgfscope}%
\end{pgfscope}%
\begin{pgfscope}%
\definecolor{textcolor}{rgb}{0.000000,0.000000,0.000000}%
\pgfsetstrokecolor{textcolor}%
\pgfsetfillcolor{textcolor}%
\pgftext[x=0.225695in, y=0.275463in, left, base]{\color{textcolor}\rmfamily\fontsize{8.000000}{9.600000}\selectfont \(\displaystyle {0}\)}%
\end{pgfscope}%
\begin{pgfscope}%
\pgfpathrectangle{\pgfqpoint{0.381946in}{0.314043in}}{\pgfqpoint{2.288996in}{1.445957in}}%
\pgfusepath{clip}%
\pgfsetbuttcap%
\pgfsetroundjoin%
\pgfsetlinewidth{0.501875pt}%
\definecolor{currentstroke}{rgb}{0.501961,0.501961,0.501961}%
\pgfsetstrokecolor{currentstroke}%
\pgfsetdash{{0.500000pt}{0.825000pt}}{0.000000pt}%
\pgfpathmoveto{\pgfqpoint{0.381946in}{0.537164in}}%
\pgfpathlineto{\pgfqpoint{2.670942in}{0.537164in}}%
\pgfusepath{stroke}%
\end{pgfscope}%
\begin{pgfscope}%
\pgfsetbuttcap%
\pgfsetroundjoin%
\definecolor{currentfill}{rgb}{0.000000,0.000000,0.000000}%
\pgfsetfillcolor{currentfill}%
\pgfsetlinewidth{0.803000pt}%
\definecolor{currentstroke}{rgb}{0.000000,0.000000,0.000000}%
\pgfsetstrokecolor{currentstroke}%
\pgfsetdash{}{0pt}%
\pgfsys@defobject{currentmarker}{\pgfqpoint{-0.048611in}{0.000000in}}{\pgfqpoint{-0.000000in}{0.000000in}}{%
\pgfpathmoveto{\pgfqpoint{-0.000000in}{0.000000in}}%
\pgfpathlineto{\pgfqpoint{-0.048611in}{0.000000in}}%
\pgfusepath{stroke,fill}%
}%
\begin{pgfscope}%
\pgfsys@transformshift{0.381946in}{0.537164in}%
\pgfsys@useobject{currentmarker}{}%
\end{pgfscope}%
\end{pgfscope}%
\begin{pgfscope}%
\definecolor{textcolor}{rgb}{0.000000,0.000000,0.000000}%
\pgfsetstrokecolor{textcolor}%
\pgfsetfillcolor{textcolor}%
\pgftext[x=0.225695in, y=0.498584in, left, base]{\color{textcolor}\rmfamily\fontsize{8.000000}{9.600000}\selectfont \(\displaystyle {2}\)}%
\end{pgfscope}%
\begin{pgfscope}%
\pgfpathrectangle{\pgfqpoint{0.381946in}{0.314043in}}{\pgfqpoint{2.288996in}{1.445957in}}%
\pgfusepath{clip}%
\pgfsetbuttcap%
\pgfsetroundjoin%
\pgfsetlinewidth{0.501875pt}%
\definecolor{currentstroke}{rgb}{0.501961,0.501961,0.501961}%
\pgfsetstrokecolor{currentstroke}%
\pgfsetdash{{0.500000pt}{0.825000pt}}{0.000000pt}%
\pgfpathmoveto{\pgfqpoint{0.381946in}{0.760285in}}%
\pgfpathlineto{\pgfqpoint{2.670942in}{0.760285in}}%
\pgfusepath{stroke}%
\end{pgfscope}%
\begin{pgfscope}%
\pgfsetbuttcap%
\pgfsetroundjoin%
\definecolor{currentfill}{rgb}{0.000000,0.000000,0.000000}%
\pgfsetfillcolor{currentfill}%
\pgfsetlinewidth{0.803000pt}%
\definecolor{currentstroke}{rgb}{0.000000,0.000000,0.000000}%
\pgfsetstrokecolor{currentstroke}%
\pgfsetdash{}{0pt}%
\pgfsys@defobject{currentmarker}{\pgfqpoint{-0.048611in}{0.000000in}}{\pgfqpoint{-0.000000in}{0.000000in}}{%
\pgfpathmoveto{\pgfqpoint{-0.000000in}{0.000000in}}%
\pgfpathlineto{\pgfqpoint{-0.048611in}{0.000000in}}%
\pgfusepath{stroke,fill}%
}%
\begin{pgfscope}%
\pgfsys@transformshift{0.381946in}{0.760285in}%
\pgfsys@useobject{currentmarker}{}%
\end{pgfscope}%
\end{pgfscope}%
\begin{pgfscope}%
\definecolor{textcolor}{rgb}{0.000000,0.000000,0.000000}%
\pgfsetstrokecolor{textcolor}%
\pgfsetfillcolor{textcolor}%
\pgftext[x=0.225695in, y=0.721705in, left, base]{\color{textcolor}\rmfamily\fontsize{8.000000}{9.600000}\selectfont \(\displaystyle {4}\)}%
\end{pgfscope}%
\begin{pgfscope}%
\pgfpathrectangle{\pgfqpoint{0.381946in}{0.314043in}}{\pgfqpoint{2.288996in}{1.445957in}}%
\pgfusepath{clip}%
\pgfsetbuttcap%
\pgfsetroundjoin%
\pgfsetlinewidth{0.501875pt}%
\definecolor{currentstroke}{rgb}{0.501961,0.501961,0.501961}%
\pgfsetstrokecolor{currentstroke}%
\pgfsetdash{{0.500000pt}{0.825000pt}}{0.000000pt}%
\pgfpathmoveto{\pgfqpoint{0.381946in}{0.983406in}}%
\pgfpathlineto{\pgfqpoint{2.670942in}{0.983406in}}%
\pgfusepath{stroke}%
\end{pgfscope}%
\begin{pgfscope}%
\pgfsetbuttcap%
\pgfsetroundjoin%
\definecolor{currentfill}{rgb}{0.000000,0.000000,0.000000}%
\pgfsetfillcolor{currentfill}%
\pgfsetlinewidth{0.803000pt}%
\definecolor{currentstroke}{rgb}{0.000000,0.000000,0.000000}%
\pgfsetstrokecolor{currentstroke}%
\pgfsetdash{}{0pt}%
\pgfsys@defobject{currentmarker}{\pgfqpoint{-0.048611in}{0.000000in}}{\pgfqpoint{-0.000000in}{0.000000in}}{%
\pgfpathmoveto{\pgfqpoint{-0.000000in}{0.000000in}}%
\pgfpathlineto{\pgfqpoint{-0.048611in}{0.000000in}}%
\pgfusepath{stroke,fill}%
}%
\begin{pgfscope}%
\pgfsys@transformshift{0.381946in}{0.983406in}%
\pgfsys@useobject{currentmarker}{}%
\end{pgfscope}%
\end{pgfscope}%
\begin{pgfscope}%
\definecolor{textcolor}{rgb}{0.000000,0.000000,0.000000}%
\pgfsetstrokecolor{textcolor}%
\pgfsetfillcolor{textcolor}%
\pgftext[x=0.225695in, y=0.944825in, left, base]{\color{textcolor}\rmfamily\fontsize{8.000000}{9.600000}\selectfont \(\displaystyle {6}\)}%
\end{pgfscope}%
\begin{pgfscope}%
\pgfpathrectangle{\pgfqpoint{0.381946in}{0.314043in}}{\pgfqpoint{2.288996in}{1.445957in}}%
\pgfusepath{clip}%
\pgfsetbuttcap%
\pgfsetroundjoin%
\pgfsetlinewidth{0.501875pt}%
\definecolor{currentstroke}{rgb}{0.501961,0.501961,0.501961}%
\pgfsetstrokecolor{currentstroke}%
\pgfsetdash{{0.500000pt}{0.825000pt}}{0.000000pt}%
\pgfpathmoveto{\pgfqpoint{0.381946in}{1.206527in}}%
\pgfpathlineto{\pgfqpoint{2.670942in}{1.206527in}}%
\pgfusepath{stroke}%
\end{pgfscope}%
\begin{pgfscope}%
\pgfsetbuttcap%
\pgfsetroundjoin%
\definecolor{currentfill}{rgb}{0.000000,0.000000,0.000000}%
\pgfsetfillcolor{currentfill}%
\pgfsetlinewidth{0.803000pt}%
\definecolor{currentstroke}{rgb}{0.000000,0.000000,0.000000}%
\pgfsetstrokecolor{currentstroke}%
\pgfsetdash{}{0pt}%
\pgfsys@defobject{currentmarker}{\pgfqpoint{-0.048611in}{0.000000in}}{\pgfqpoint{-0.000000in}{0.000000in}}{%
\pgfpathmoveto{\pgfqpoint{-0.000000in}{0.000000in}}%
\pgfpathlineto{\pgfqpoint{-0.048611in}{0.000000in}}%
\pgfusepath{stroke,fill}%
}%
\begin{pgfscope}%
\pgfsys@transformshift{0.381946in}{1.206527in}%
\pgfsys@useobject{currentmarker}{}%
\end{pgfscope}%
\end{pgfscope}%
\begin{pgfscope}%
\definecolor{textcolor}{rgb}{0.000000,0.000000,0.000000}%
\pgfsetstrokecolor{textcolor}%
\pgfsetfillcolor{textcolor}%
\pgftext[x=0.225695in, y=1.167946in, left, base]{\color{textcolor}\rmfamily\fontsize{8.000000}{9.600000}\selectfont \(\displaystyle {8}\)}%
\end{pgfscope}%
\begin{pgfscope}%
\pgfpathrectangle{\pgfqpoint{0.381946in}{0.314043in}}{\pgfqpoint{2.288996in}{1.445957in}}%
\pgfusepath{clip}%
\pgfsetbuttcap%
\pgfsetroundjoin%
\pgfsetlinewidth{0.501875pt}%
\definecolor{currentstroke}{rgb}{0.501961,0.501961,0.501961}%
\pgfsetstrokecolor{currentstroke}%
\pgfsetdash{{0.500000pt}{0.825000pt}}{0.000000pt}%
\pgfpathmoveto{\pgfqpoint{0.381946in}{1.429647in}}%
\pgfpathlineto{\pgfqpoint{2.670942in}{1.429647in}}%
\pgfusepath{stroke}%
\end{pgfscope}%
\begin{pgfscope}%
\pgfsetbuttcap%
\pgfsetroundjoin%
\definecolor{currentfill}{rgb}{0.000000,0.000000,0.000000}%
\pgfsetfillcolor{currentfill}%
\pgfsetlinewidth{0.803000pt}%
\definecolor{currentstroke}{rgb}{0.000000,0.000000,0.000000}%
\pgfsetstrokecolor{currentstroke}%
\pgfsetdash{}{0pt}%
\pgfsys@defobject{currentmarker}{\pgfqpoint{-0.048611in}{0.000000in}}{\pgfqpoint{-0.000000in}{0.000000in}}{%
\pgfpathmoveto{\pgfqpoint{-0.000000in}{0.000000in}}%
\pgfpathlineto{\pgfqpoint{-0.048611in}{0.000000in}}%
\pgfusepath{stroke,fill}%
}%
\begin{pgfscope}%
\pgfsys@transformshift{0.381946in}{1.429647in}%
\pgfsys@useobject{currentmarker}{}%
\end{pgfscope}%
\end{pgfscope}%
\begin{pgfscope}%
\definecolor{textcolor}{rgb}{0.000000,0.000000,0.000000}%
\pgfsetstrokecolor{textcolor}%
\pgfsetfillcolor{textcolor}%
\pgftext[x=0.166667in, y=1.391067in, left, base]{\color{textcolor}\rmfamily\fontsize{8.000000}{9.600000}\selectfont \(\displaystyle {10}\)}%
\end{pgfscope}%
\begin{pgfscope}%
\pgfpathrectangle{\pgfqpoint{0.381946in}{0.314043in}}{\pgfqpoint{2.288996in}{1.445957in}}%
\pgfusepath{clip}%
\pgfsetbuttcap%
\pgfsetroundjoin%
\pgfsetlinewidth{0.501875pt}%
\definecolor{currentstroke}{rgb}{0.501961,0.501961,0.501961}%
\pgfsetstrokecolor{currentstroke}%
\pgfsetdash{{0.500000pt}{0.825000pt}}{0.000000pt}%
\pgfpathmoveto{\pgfqpoint{0.381946in}{1.652768in}}%
\pgfpathlineto{\pgfqpoint{2.670942in}{1.652768in}}%
\pgfusepath{stroke}%
\end{pgfscope}%
\begin{pgfscope}%
\pgfsetbuttcap%
\pgfsetroundjoin%
\definecolor{currentfill}{rgb}{0.000000,0.000000,0.000000}%
\pgfsetfillcolor{currentfill}%
\pgfsetlinewidth{0.803000pt}%
\definecolor{currentstroke}{rgb}{0.000000,0.000000,0.000000}%
\pgfsetstrokecolor{currentstroke}%
\pgfsetdash{}{0pt}%
\pgfsys@defobject{currentmarker}{\pgfqpoint{-0.048611in}{0.000000in}}{\pgfqpoint{-0.000000in}{0.000000in}}{%
\pgfpathmoveto{\pgfqpoint{-0.000000in}{0.000000in}}%
\pgfpathlineto{\pgfqpoint{-0.048611in}{0.000000in}}%
\pgfusepath{stroke,fill}%
}%
\begin{pgfscope}%
\pgfsys@transformshift{0.381946in}{1.652768in}%
\pgfsys@useobject{currentmarker}{}%
\end{pgfscope}%
\end{pgfscope}%
\begin{pgfscope}%
\definecolor{textcolor}{rgb}{0.000000,0.000000,0.000000}%
\pgfsetstrokecolor{textcolor}%
\pgfsetfillcolor{textcolor}%
\pgftext[x=0.166667in, y=1.614188in, left, base]{\color{textcolor}\rmfamily\fontsize{8.000000}{9.600000}\selectfont \(\displaystyle {12}\)}%
\end{pgfscope}%
\begin{pgfscope}%
\definecolor{textcolor}{rgb}{0.000000,0.000000,0.000000}%
\pgfsetstrokecolor{textcolor}%
\pgfsetfillcolor{textcolor}%
\pgftext[x=0.111111in,y=1.037022in,,bottom,rotate=90.000000]{\color{textcolor}\rmfamily\fontsize{8.000000}{9.600000}\selectfont time [µs]}%
\end{pgfscope}%
\begin{pgfscope}%
\pgfsetrectcap%
\pgfsetmiterjoin%
\pgfsetlinewidth{0.803000pt}%
\definecolor{currentstroke}{rgb}{0.000000,0.000000,0.000000}%
\pgfsetstrokecolor{currentstroke}%
\pgfsetdash{}{0pt}%
\pgfpathmoveto{\pgfqpoint{0.381946in}{0.314043in}}%
\pgfpathlineto{\pgfqpoint{0.381946in}{1.760000in}}%
\pgfusepath{stroke}%
\end{pgfscope}%
\begin{pgfscope}%
\pgfsetrectcap%
\pgfsetmiterjoin%
\pgfsetlinewidth{0.803000pt}%
\definecolor{currentstroke}{rgb}{0.000000,0.000000,0.000000}%
\pgfsetstrokecolor{currentstroke}%
\pgfsetdash{}{0pt}%
\pgfpathmoveto{\pgfqpoint{2.670942in}{0.314043in}}%
\pgfpathlineto{\pgfqpoint{2.670942in}{1.760000in}}%
\pgfusepath{stroke}%
\end{pgfscope}%
\begin{pgfscope}%
\pgfsetrectcap%
\pgfsetmiterjoin%
\pgfsetlinewidth{0.803000pt}%
\definecolor{currentstroke}{rgb}{0.000000,0.000000,0.000000}%
\pgfsetstrokecolor{currentstroke}%
\pgfsetdash{}{0pt}%
\pgfpathmoveto{\pgfqpoint{0.381946in}{0.314043in}}%
\pgfpathlineto{\pgfqpoint{2.670942in}{0.314043in}}%
\pgfusepath{stroke}%
\end{pgfscope}%
\begin{pgfscope}%
\pgfsetrectcap%
\pgfsetmiterjoin%
\pgfsetlinewidth{0.803000pt}%
\definecolor{currentstroke}{rgb}{0.000000,0.000000,0.000000}%
\pgfsetstrokecolor{currentstroke}%
\pgfsetdash{}{0pt}%
\pgfpathmoveto{\pgfqpoint{0.381946in}{1.760000in}}%
\pgfpathlineto{\pgfqpoint{2.670942in}{1.760000in}}%
\pgfusepath{stroke}%
\end{pgfscope}%
\end{pgfpicture}%
\makeatother%
\endgroup%